%% file: tauetapipaperv2rev.tex
\documentclass[dvips,a4paper,12pt]{article}
\usepackage{graphicx}
\usepackage{amsmath}
\usepackage{authblk}
\usepackage[numbers,sort&compress]{natbib}
\usepackage[bookmarks=true,colorlinks=true,linkcolor=black,citecolor=magenta]{hyperref}

\input{newcommands}

\newcommand{\lambetapi}[1]{\sqrt{\lambda_{\eta\pi}{(#1)}}}
\newcommand{\lambetapid}[1]{\lambda_{\eta\pi}{(#1)}}

\usepackage{calc}
\newcommand{\styletofont}[1]{%
  \ifx\displaystyle#1\let\next\textfont\fi
  \ifx\textstyle#1\let\next\textfont\fi
  \ifx\scriptstyle#1\let\next\scriptfont\fi
  \ifx\scriptscriptstyle#1\let\next\scriptscriptfont\fi}

\newcommand{\innfwhat}[2]{%
  \styletofont{#1}%
  \dimen0 \fontcharic\next1 \skewchar\next1
  \advance\dimen0 -\fontcharic\next1`#2%
  \makebox[0pt][l]{$#1#2$}%
  \makebox[\widthof{$#1#2$}]{$#1\kern.5\dimen0 \widehat{\vphantom{#2}}$}}

\title{\vskip-3cm\hfill{\small LPT-ORSAY/14-17}\\[2cm]
Analyticity of $\eta\pi$ isospin-violating form factors and
  the $\tau\to\eta\pi\nu$ second-class decay}

\author[a]{S. Descotes-Genon}

\affil[a]{\small Laboratoire de Physique Th\'eorique, 
  CNRS/Univ. Paris-Sud 11 (UMR8627), 91405~Orsay, France}

\author[b]{B. Moussallam}

\affil[b]{\small Groupe de Physique Th\'eorique,
    IPN, CNRS/IN2P3/Univ. Paris-Sud 11 (UMR8608), 91406 Orsay, France}

\begin{document}

\date{\today}

\maketitle

\begin{abstract}
We consider the evaluation of the $\eta\pi$ isospin-violating
vector and scalar form factors relying on a systematic application of
analyticity and unitarity, combined with chiral expansion results. 
It is argued that the usual analyticity properties do hold (i.e. no
anomalous thresholds are present) in spite of the instability of the
$\eta$ meson in QCD. 
Unitarity relates the vector form factor to the $\eta\pi \to \pi\pi$ 
amplitude: we exploit progress in formulating and solving the
Khuri-Treiman equations for $\eta\to 3\pi$ and in experimental
measurements of the Dalitz plot parameters to evaluate the shape
of the $\rho$-meson peak.
Observing  this peak in the energy distribution of the $\tau\to
\eta \pi \nu$ decay would be a background-free signature of a 
second-class amplitude.  
The scalar form factor is also estimated from a phase dispersive
representation using a  plausible model  for the $\eta\pi$ elastic
scattering $S$-wave phase shift and a sum rule constraint in the
inelastic region. We indicate how a possibly exotic nature of the
$a_0(980)$ scalar meson manifests itself in a dispersive approach. A
remark is finally made on a second-class amplitude in the
$\tau\to\pi\pi\nu$ decay. 

\end{abstract}

\tableofcontents

\section{Introduction}
Isospin-breaking phenomena involving light pseudoscalar mesons are
particularly interesting probes of the  three flavour chiral expansion
as they are driven by the parameter
\be\lbl{epsilon}
\epsilon={\sqrt3(m_d-m_u)\over4(m_s-(m_u+m_d)/2) }
\en
which involves the three light-quark masses. 
One of the major goals of non-perturbative approaches to QCD is to
arrive at an accurate determination of the light-quark masses. One 
issue in the direct determination of $\epsilon$ from  the very
precisely known difference between the masses of the charged and the
neutral kaon is to properly evaluate the electromagnetic contribution
to this difference. At leading chiral order, it is given by Dashen's 
low-energy theorem~\cite{Dashen:1969eg}. There are, however, possible
substantial corrections from next-to-leading $O(e^2 m_s)$ effects
which suggests to explore, in parallel, other isospin-violating
processes. 

In this respect, the $\eta\to 3\pi$ decay
amplitude is of particular interest since electromagnetic
contributions are absent at leading order~\cite{Sutherland:1966zz} and
found to be rather small at next-to-leading order  
(NLO)~\cite{Baur:1995gc,Deandrea:2008px,Ditsche:2008cq}. 
Furthermore, there has been considerable progress, on the experimental
side, in the precision of the measurements of the Dalitz plot
parameters for both the $\eta\to 3\pi^0$
amplitude\cite{Adolph:2008vn,Unverzagt:2008ny,Prakhov:2008ff,Ambrosinod:2010mj}
and the $\eta\to\pi^+\pi^-\pi^0$ amplitude\cite{Ambrosino:2008ht}.
There is a price to pay, however, in that the chiral expansion
has an inherently slow convergence in the treatment of final-state 
interactions, e.g. in the NLO expression of the $\eta\to 3\pi$
amplitude~\cite{Gasser:1984pr}, $\pi\pi$ rescattering is treated only
at leading chiral order. The amplitude has now been computed to NNLO
in chiral perturbation theory (ChPT)~\cite{Bijnens:2007pr}. A partly
analytic representation  of the $\pi\pi$ rescattering part at NNLO,
accounting for some effects of higher order, was obtained in
ref.~\cite{Kampf:2011wr}.  A treatment of rescattering in the
framework of non-relativistic effective field theory has been
discussed in ref.~\cite{Schneider:2010hs}.  
An alternative approach is to combine the chiral expansion with a more
general representation which encodes exact unitarity, analyticity and
crossing symmetry~\cite{Roiesnel:1980gd}. A rigorous framework was
proposed by Khuri and Treiman (KT)~\cite{Khuri:1960zz} who derived a
set of integral equations for the analogous $K\to 3\pi$
problem. Application to the $\eta\to3\pi$ amplitude was first
discussed in ref.~\cite{Neveu:1970tn}.  The KT equations were more
recently generalised to account for both $S$ and $P$-wave elastic
rescattering~\cite{Kambor:1995yc,Anisovich:1996tx} and numerical
solutions were constructed.  Updates of these analyses, which take
into account the recent experimental data, have been
presented~\cite{Colangelo:2011zz,Lanz:2013ku,Leutwyler:2013wna}.

This progress have motivated us to reconsider the problem of
evaluating the $\eta\pi$ isospin-violating vector and scalar form
factors exploiting, as systematically as possible, their analyticity
properties and matching with chiral NLO
calculations~\cite{Neufeld:1994eg,Scora:1995sj}. The vector form
factor, in particular, probes the $\eta\pi\to \pi\pi$ amplitude, via
unitarity, in a kinematical region different from that of 
the decay, but where the KT equations should still be applicable. 
These two form factors are measurable, in principle, from the $\tau\to
\eta\pi\nu$ decay mode. This mode, being forbidden in the 
isospin-symmetric limit, is a clean example of  the ``second-class
currents'' as introduced by Weinberg~\cite{Weinberg:1958ut}, which are
yet to be discovered experimentally. An upper bound on the branching
fraction, $B_{\eta\pi} <  1.4\times 10^{-4}$,  was obtained by the CLEO
collaboration\cite{Bartelt:1996iv} which was  slightly improved to
$B_{\eta\pi} <  9.9\times 10^{-5}$ by Babar~\cite{delAmoSanchez:2010pc}. 
The Belle collaboration has quoted $B_{\eta\pi} <  7.3\times 10^{-5}$ as
a preliminary result~\cite{Hayasaka:2009zz} which, however, was not
subsequently confirmed. 
Theoretical estimates for
this branching fraction~\cite{Tisserant:1982fc,Pich:1987qq,Bednyakov:1992af,Neufeld:1994eg,Nussinov:2008gx,Paver:2010mz,Volkov:2012be} yield values in the range
$10^{-6}$ to $10^{-5}$, which do not seem so small as compared to the
number of $\tau$ pairs available at Babar: $N_\tau\simeq 4.3\times
10^8$ or Belle $N_\tau\simeq 9.0\times 10^8$.  
One difficulty faced by the $B$ factories was that a
substantial number of $\eta\pi$ pairs  were produced from background
modes, like $\tau^-\to \eta\pi^-\pi^0\nu$, which have to be subtracted. 
A drastic reduction of this background should be
possible at $\tau$-charm factories~\cite{Asner:2008nq,Bondar:2013cja}
which could make detailed measurements of the $\eta\pi$ mode possible.
An increase in the luminosity by a factor of fifty is expected at
Belle II~\cite{Akeroyd:2004mj}.

Here, we will consider not only the integrated $\tau\to
\eta\pi\nu_\tau$ branching  fraction but also the detailed 
dependence as a function of the $\eta\pi$ invariant mass. 
This dependence carries nontrivial dynamical
information. Close to $s=0$ it can be related, via ChPT, to the 
isospin-breaking quark mass ratio $\epsilon$ while, at higher energy, 
the shape of the $\rho$ resonance peak in the vector form factor can
be related to $\eta\to 3\pi$ decay properties. From an experimental
point of view, the observation of the $\rho$ peak would be a
background-free signal of a second-class amplitude. A peak at the
$a_0(980)$ mass is also expected from the scalar form factor. As was
noted long ago~\cite{Bramon:1987zb}, the mode $\tau\to \eta\pi \nu$
probes the ``nature'' of the scalar meson $a_0(980)$ in a clean way
via its coupling to the $\bar{u}d$ operator. 

The $\eta\pi$ scalar form factor has a potential for constraining
extensions of the Standard Model which contain charged Higgs
bosons. For illustration, in the two-Higgs model proposed in
ref.~\cite{Pich:2009sp}\footnote{This model, in which tree-level
  flavour-changing neutral currents are avoided by an alignment
  prescription of Yukawa matrices, includes a number of previously
  proposed models and also allows for CP violation.}
, the energy dependence of the form factor is
modified as follows,
\be
\fzero(s)=\left. \fzero\right(s)\vert_{SM}\left( 
1-{\zeta^*_\tau (\zeta_u m_u-\zeta_d m_d)\over m_u-m_d}\times
{s\over m^2_{H^+}}  \right)
\en
where the $\zeta$'s are coupling constants. The influence of the
charged Higgs in this form factor could be enhanced because of the
$m_u-m_d$ denominator, depending on the relative sign and size of
$\zeta_u$, $\zeta_d$. The constraints which are already available
(specifically from $B\to\tau \nu$) are not so stringent:
$\vert\zeta_l\zeta_d/m^2_{H^+}\vert < 0.1$ 
$\hbox{GeV}^{-2}$~\cite{Jung:2010ik}. In order to  derive a similar
level of constraint, one should be able to evaluate 
$\fzero$ in the Standard Model (and also be able to measure it, of
course) with a precision of $\simeq 20\%$ at $s=1\ \hbox{GeV}^2$. 

The plan of the paper is as follows. After introducing some basic
formulae and notation, we list the contributions from the light
two-meson states to the unitarity relations of the two form
factors. We also discuss contributions with one photon. Then, we
recall the main results from the NLO ChPT
calculations~\cite{Neufeld:1994eg,Scora:1995sj}: the values of the
form factors at $s=0$ and their first derivatives will be used as
input in the dispersive representations. In order to derive these, it
is important to check the possible presence of anomalous thresholds,
since the $\eta$ meson is unstable: we present arguments that they are
actually absent. We then discuss the dispersive evaluation of the
vector form factor using as input $\pi\eta\to \pi\pi$ amplitudes
satisfying the KT equations and constrained by experimental data in
the physical decay region. Finally, we estimate the scalar form factor
from a phase dispersive representation, using a modelling of
$\eta\pi\to \eta\pi$ elastic scattering borrowed from
ref.~\cite{Black:1999dx}.

\section{Definitions and basic unitarity relations}\lblsec{defsbasic}
The semi-leptonic weak decay amplitudes $\tau\to\eta\pi\nu$ and
$\eta\to l\pi\nu$ (with $l=e,\mu$) are induced by the usual Fermi
Lagrangian 
\be\lbl{fermilag}
{\cal L}_{F}=-{G_F V_{ud}\over\sqrt2}\left[
  \bar{u}\gamma^\mu(1-\gamma^5) d\times 
\bar{l} \gamma_\mu(1-\gamma^5) \nu_l + h.c.\right]\ .
\en 
The $\eta\pi$ matrix element of the charged vector current is expressed
in terms of two form factors (we follow the same notation as
ref.~\cite{Neufeld:1994eg} except that we call the $\eta\pi$ invariant
mass squared $s$ instead of $t$), 
\be\lbl{basiquedef}
\outbraque{\eta(p_\eta)\pi^+(p_\pi)\vert \Vmud(0) \vert 0}=
-\sqrt2\left[ f_+^{\eta\pi}(s)(p_\eta-p_\piplus)_\mu +
  f_-^{\eta\pi}(s)(p_\eta+p_\piplus)_\mu\right]\  
\en
with 
\be
\Vmud(x)=\bar{u}(x)\gamma_\mu d(x),\quad s=(p_\eta+p_\pi)^2\ . 
\en
When writing unitarity relations it is convenient to introduce the
scalar form factor $f_0^{\eta\pi}(s)$ instead of $f_-^{\eta\pi}(s)$
\be
f_0^{\eta\pi}(s)= f_+^{\eta\pi}(s) +{s\over\Delta_{\eta\pi}}
f_-^{\eta\pi}(s),\quad
\Delta_{PQ}=m_P^2-m_Q^2
\en
The expression for the differential decay width of the $\tau$ lepton
which derives from the Fermi Lagrangian~\rf{fermilag} and the
definition of the form factors~\rf{basiquedef} then reads
\bea\lbl{differtaudecay}
&& {d\Gamma_{\tau\to\eta\pi\nu_\tau}\over ds}={G_F^2 V_{ud}^2
  S_{EW}\, m_\tau^3 \over 384\,\pi^3 }{\lambetapi{s}\over
  s^3} \left(1-{s\over\mtaud}\right)^2\nonumber\\
&& \quad\quad \times \bigg\{ \vert\fplus(s)\vert^2\,\lambetapid{s}
   \left(1+{2s\over\mtaud}\right)  
+3 \vert\fzero(s) \vert^2 \Delta_{\eta\pi}^2 \bigg\}
\ena 
%modif: S_{EW} defined
where $S_{EW}$ is the logarithmically enhanced universal radiative
correction factor~\cite{Sirlin:1977sv}
($S_{EW}=1.0201$~\cite{Erler:2002mv})
and
\be
\lambda_{PQ}(s)=\lambda(s,m_P^2,m_Q^2)\ ,
\en
$\lambda$ being the K\"all\'en function
$\lambda(x,y,z)=x^2+y^2+z^2-2(xy+yz+xz)$. 
For the physical
$\tau$ decay, the variable $s$ lies in the range $(\meta+\mpi)^2\le
s\le \mtaud$.   
The analogous formula for the differential decay of the $\eta$,
$\eta\to l^\pm \pi^\mp \nu_l$ with $l=e$ or $l=\mu$ reads
\bea
&& {d\Gamma_{\eta\to l \pi \nu_l}\over ds}={G_F^2 V_{ud}^2 S_{EW}\over
  192 \pi^3 } {\lambetapi{s}\over m_\eta^3}  
\left( 1-{m_l^2\over s}\right)^2         \nonumber\\
&&  \quad\quad\times\left\{ 
 \vert\fplus(s)\vert^2\,\lambetapid{s}\left(2+{m_l^2\over s}\right) 
+3 \vert\fzero(s)\vert^2 \Delta_{\eta\pi}^2 {m_l^2\over s}
\right\}\ .
\ena
In this case, the variable $s$ is restricted to the range 
$ m_l^2\le s\le (\meta-\mpi)^2$.
\subsection{Unitarity relations for $\fplus(s)$ }\lblsec{unitfplus}
We consider the $\eta\pi^+$ center-of-mass system and choose the
$z$-axis along the three-momentum of the $\eta$ meson.
The vector form factor is easily seen to be proportional to the  
matrix element of the third component of the vector current in this frame
\be
\outbraque{\eta(p_\eta)\pi^+(p_\pi)\vert j_3^{ud}(0)\vert 0}=
-2\sqrt2 q_{\eta\pi}(s)\,\fplus(s)\ ,
\en
where $q_{\eta\pi}(s)$ is the center-of-mass momentum. 
The form factor $\fplus(s)$ can be defined as an analytic function of $s$
with a cut along the positive real axis starting at $s_{th}=4\mpid$
(see the discussion about the absence of anomalous thresholds in
sec.~\sect{remarkanomth} below).
The discontinuity across the cut has
the form of a generalised unitarity relation and is given as a sum
over a complete set of states,
\be\lbl{discfplus}
-2\sqrt2\,q_{\eta\pi}(s)\,\disc[\fplus(s)]=
{1\over2}\sum_n T^*_{\eta\pi^+\to n} \times
\outbraque{n\vert j_3^{ud}(0) \vert 0} \ .
\en
with
\be
\disc[\fplus(s)]\equiv {\fplus(s+i\epsilon)-\fplus(s-i\epsilon)
  \over2i}\ .
\en
%modif: l=1 -> angular momentum l=1
The lightest state contributing to the unitarity relation is
$n=\pi^0\pi^+$ with angular momentum $l=1$. The next-to-lightest
contribution is from four pion states $n=\pi^0\pi^+\pi\pi$. However,
we expect such contributions not to be effectively relevant below 1
GeV because of phase space suppression and we will ignore them here.
Let us consider successively the contributions from the lightest
two-body states $n=\pi^0\pi^+$, $n=\eta\pi^+$ and $n=\bar{K}^0 K^+$
\begin{itemize}
\item[{\bf a)}] $n=\pi^0\pi^+$:\\
The $\pi^0\pi^+$ matrix element of the vector current,
\be\lbl{Fvpidef}
\outbraque{\pi^0\pi^+\vert\Vmud\vert0}=\sqrt2
\left[ F_V^\pi(s)(p_{\pizero}-p_{\piplus})_\mu
     + F_-^\pi(s)(p_{\pizero}+p_{\piplus})_\mu \right]\ ,
\en
involves two form factors  since $\mpiplus\ne\mpizero$. The unitarity
relation for $f_+^{\eta\pi}$ involves only the vector  form factor,
$F_V^\pi(s)$.  
Using eq.~\rf{discfplus}, we can write the
$\pi\pi$ contribution in the unitarity relation as follows   
\bea\lbl{unitfplus}
&& \disc\left[\fplus(s)\right]_{\pi\pi}=- \theta(s-4\mpid) {s-4\mpid\over
    16\pi\, \lambetapi{s} }\, F_V^\pi(s)\nonumber\\ 
&& \phantom{\left.\disc\fplus(s)\right\vert_{\pi\pi}}
\quad\times {1\over2}\int_{-1}^1 dzz\, T^*_{\pi^0\pi^+\to \eta\pi^+}(s,t,u)\ ,
\ena
where $z=\cos\theta$, $\theta$ being the scattering angle in the
center-of-mass  system. We expect this contribution to be important
below 1 GeV, because of the  presence  of the $\rho(770)$ resonance.
\item[{\bf b)}] $n=\eta\pi^+$\\
Next, the contribution from the $\eta\pi$ state to  the unitarity
relation reads
\bea\lbl{etapipwave}
&& \disc\left[\fplus(s)\right]_{\eta\pi}=\theta(s-(m_\eta+m_\pi)^2)
{\lambetapi{s}\over16\pi  s} \fplus(s)
\nonumber\\
&& \phantom{ \disc\left[\fplus(s)\right]_{\eta\pi}}
\times{1\over2}\int_{-1}^1 dz\, z\, T^*_{\eta\pi^+\to\eta\pi^+}(s,t,u)
\ena
This contribution involves the $\eta\pi\to\eta\pi$ amplitude projected
on the $P$-wave. The quantum numbers of the state $(\eta\pi)_{l=1}$ are exotic:
$J^{PC}=1^{-+}$. We expect the $(\eta\pi\to\eta\pi)_{l=1}$ amplitude to
be very small below 1 GeV. This is borne out by the $\eta\pi$
scattering model proposed in ref.~\cite{Black:1999dx}, which 
predicts that the $P$-wave phase shift is of the order of $-1^\circ$
at 1 GeV.

\item[{\bf c)}] $n=\kzerob\kplus$\\
Finally, let us write the contribution of the $\kzerob\kplus$ state in the
unitarity relation, which is useful for comparing with the chiral
calculation. In this case, the kaon vector form factor appears,
defined from
\be\lbl{FVKdef}
\outbraque{\KzeroK \vert  \Vmud \vert 0}=-\left[
F_V^K(s) (p_{\kzerob}-p_{\kplus})_\mu +F_-^K(s)
(p_{\kzerob}+p_{\kplus})_\mu\right]\ .
\en
The corresponding contribution in the unitarity relation reads
\bea\lbl{unitfplusKK}
&& \disc\left[\fplus(s)\right]_{\bar{K}K}= \theta(s-4\mkd) {s-4\mkd\over
16\pi\sqrt2\, \lambetapi{s}}
\, F_V^K(s) \nonumber\\
&& \phantom{\disc\left[\fplus(s)\right]_{\bar{K}K}}
 \times{1\over2}\int_{-1}^1 dzz T^*_{\KzeroK\to \eta\pi^+}(s,t,u)\ .
\ena
In the above expression,  isospin breaking is contained in the
amplitude ${\KzeroK\to \eta\pi^+}$ projected on the $P$-wave. 
Let us recall the reason: since
$G\ket{K^+}=\ket{\bar{K}_0}$, $G\ket{\bar{K}_0}=-\ket{K^+}$ 
one has $G\vert\bar{K}_0K^+\rangle= (-1)^{l+1} \vert \bar{K}_0K^+\rangle$. 
Since $\eta\pi$ has G-parity $-1$, this implies that the 
partial-wave amplitudes $(\KzeroK\to \eta\pi^+)_l$ with odd 
angular momentum $l$ vanish in the isospin limit. 
\end{itemize}

\subsection{Unitarity relations for $\fzero(s)$}\lblsec{unitfzero}
Unitarity relations for the scalar form factor can be derived in exactly
the same way as above noticing that, in the center-of-mass frame, the
matrix element of the zeroth component of the vector current is
proportional to $\fzero$,
\be
\outbraque{\eta(p_\eta)\pi^+(p_\pi)\vert j^{ud}_0(0) \vert
  0}=-{\sqrt2\Delta_{\eta\pi} \over \sqrt{s}}\,f_0^{\eta\pi}(s)
\en
One can then derive a relation for the discontinuity along the cut,
analogous to eq.~\rf{discfplus},
\be\lbl{discfzero}
-{\sqrt2\Delta_{\eta\pi} \over \sqrt{s}}\,\disc[f_0^{\eta\pi}(s)]=
{1\over2}\sum_n T^*_{\eta\pi^+\to n} \times
\outbraque{n\vert j_0^{ud}(0) \vert 0} \ .
\en
As before, let us consider the contributions from the lightest
two-particle states  $\pi\pi$, $\pi\eta$ and $K\Kbar$.
\begin{itemize}
\item[{\bf a)}] $n=\pi^0\pi^+$:\\
Introducing a scalar pion form factor\footnote{This form
  factor induces a second-class amplitude in the $\tau^\pm\to
  \pi^0\pi^\pm\nu$ decay, see Appendix~\sect{taupipi2nd}}
  from eq.~\rf{Fvpidef} 
\be\lbl{f0pipi}
f_0^{\pi\pi}(s)=F_V^\pi(s)+{s\over\Delta_{\pi^0\pi^+}} F_-^\pi(s)
\en
one derives that
\bea\lbl{discf0pipi}
&& \disc\left[ \fzero\right]_{\pi^0\pi^+}=-\theta(s-4\mpid)
{\sqrt{s-4\mpid}\over16\pi\sqrt{s}}{\Delta_{\pi^0\pi^+}\over \Delta_{\eta\pi^+}}
f_0^{\pi\pi}(s)\nonumber\\
&&\phantom{ \disc\left[ \fzero\right]_{\pi^0\pi^+}}
\times {1\over2}\int_{-1}^1 dz\,T^*_{\eta\pi^+  \to  \pi^0\pi^+}(s,t,u)\ .
\ena
This contribution involves a product of two isospin-breaking terms
($\Delta_{\pi^0\pi^+}$ and $T^*_{\eta\pi^+\to\pi^0\pi^+}$) and thus
must be negligibly small in practice.
\item[{\bf b)}] $n=\eta\pi^+$:\\
The contribution from the $\eta\pi^+$ states to the unitarity relation
reads 
\bea
&&\disc\left[\fzero\right]_{\eta\pi}= \theta(s-(\meta+\mpi)^2))
{\lambetapi{s}\over16\pi s}\fzero(s)
\nonumber\\
&&\phantom{\disc\left[\fzero\right]_{\eta\pi}}
\times{1\over2}\int_{-1}^1 dz\, T^*_{\eta\pi^+\to\eta\pi^+}(s,t,u)\ .
\ena
It has a form similar to eq.~\rf{etapipwave} for the vector form
factor except that it involves the $\eta\pi\to \eta\pi$ amplitude
projected on the $S$-wave instead of the $P$-wave. This contribution
is enhanced by the presence of the $a_0(980)$ resonance and thus must
be the dominating one below 1 GeV.

\item[{\bf c)}] $n=\kzerob\kplus$:\\
Finally, the contribution from $\kzerob\kplus$ involves the
corresponding scalar form factor
\be
f_0^{\kzerob\kplus}(s)\equiv F_V^K(s) + {s\over \Delta_{\kzerob\kplus}}F_-^K(s)
\en
and it has the following expression
\bea\lbl{discfzeroKK}
&& \disc\left[\fzero\right]_{\kzerob\kplus}(s)=\theta(s-4\mkd)
{\sqrt{s-4\mkd}\over16\pi\sqrt{s}}{\Delta_{\kzerob\kplus}\over\sqrt2\Delta_{\eta\pi}
} f_0^{\kzerob\kplus}(s)\nonumber\\
&&\phantom{\disc\left[\fzero\right]_{\kzerob\kplus}(s)}
\times{1\over2}\int_{-1}^1 dz\, T^*_{\eta\pi^+\to \kzerob\kplus}(s,t,u)\ .
\ena
As compared to the analogous contribution for $\fplus$, the
relation~\rf{discfzeroKK} involves the $\eta\pi^+\to \kzerob\kplus$
amplitude projected on the $S$-wave, which is isospin
conserving. Isospin breaking is contained in the mass difference
factor $\Delta_{\kzerob\kplus}$.
\end{itemize}
\subsection{Some electromagnetic contributions to the unitarity relations}
In the unitarity equations discussed above, we have considered only hadronic
states in the sums over $n$. Since we are studying isospin-breaking form
factors, electromagnetic contributions are present and, at order $e^2$ one
should also consider states involving one photon.  Note that EM contributions
have already appeared, e.g. in eqs.~\rf{discf0pipi},~\rf{discfzeroKK} which
are proportional to the mass differences $m_\pizero^2-m_\piplus^2$ (which is
mainly electromagnetic) and $m_\kzerob^2-m_\kplus^2$ (which is partly of
electromagnetic origin). These contributions are dominant in the chiral
counting; they are included in the NLO chiral expressions.
We will not discuss EM contributions in their full generality here and simply
mention the contributions of the two lightest states $n=\gamma\pi$ and
$n=\gamma\pi\pi$  in the unitarity relations:
\begin{itemize}
\item[{\bf a)}]   $n=\gamma\pi$ \\
The $\gamma\pi$ matrix element of the vector current can be expressed
in terms of one form factor
\be\lbl{gammpiff}
\braque{\gamma(\lambda)\pi^+\vert\Vmud(0)\vert
  0}=eF_V^{\pi\gamma}(s)\,\varepsilon_\mu[e_\gamma(\lambda),p_\gamma,p_\piplus] 
\en
where $e_\gamma$ is the polarisation vector of the photon\footnote{We
use the  simplified notation: $\varepsilon_\mu(a,b,c)\equiv
  \varepsilon_{\mu\nu\alpha\beta}\, a^\nu b^\alpha c^\beta $ and the
  convention $\varepsilon_{0123}=+1$.}. 
At leading order in the chiral expansion, the value of this form
factor at $s=0$ is given by the anomaly
\be
\left.F_V^{\gamma\pi}(0)\right\vert_{LO}={\sqrt2 N_c\over 24\pi^2
  \fpid}\ .
\en
Going to the center-of-mass frame, one sees that the matrix
element~\rf{gammpiff} vanishes for $j_0^{\eta\pi}$.  The unitarity
contribution from $\gamma\pi$  thus concerns only the 
vector form factor.  One finds the following expression for the
discontinuity: 
\bea
&&\disc[\fplus(s)]_{\pi\gamma}=\theta(s-\mpid)
{i(s-\mpid)^2\over128\pi\,s\,q_{\eta\pi}(s)}
e F_V^{\pi\gamma}(s)\nonumber\\
&&\phantom{\disc[\fplus(s)]_{\pi\gamma}}
\times{1\over2}\sum_{\lambda=\pm1}\int_0^\pi
d\theta\sin^2\theta\,T^*_{\eta\piplus\to\gamma(\lambda)\piplus}\ . 
\ena
Evaluating this contribution precisely would require some modelling of
the amplitude $\eta\pi\to\gamma\pi$. It is likely that this amplitude
should be small below 1 GeV since no resonant contribution from the
$\rho$-meson is allowed in the isospin limit.

\item[{\bf b)}]   $n=\gamma\pi\pi$ \\
In principle, the states $n=\gamma\pi\pi$ can contribute to the
unitarity relations for both $\fplus$ and $\fzero$. We will consider
here only the latter one, the evaluation of which is simplified by
using the relation between $\fzero$ and the matrix element of the
divergence,
\be
\outbraque{\eta(p_\eta)\pi(p_\pi)\vert i\partial^\mu \Vmud(0)\vert 0}=
\sqrt2\Delta_{\eta\pi}\,\fzero(s)\ ,
\en
together with the Ward identity for the divergence,
\be\lbl{wardidentity}
i\partial^\mu \Vmud(x)=(m_d-m_u)\bar{u}d(x)-eA^\mu(x)\Vmud(x)\ .
\en
Eq.~\rf{wardidentity} makes it easy to evaluate the matrix element
involving $\gamma\pi\pi$ in terms of the pion vector form factor
\be
\outbraque{\gamma(\lambda)\pizero\piplus\vert \partial^\mu
  \Vmud(0)\vert0}=-e\sqrt2\,e_\gamma(\lambda)\cdot (p_\pizero-p_\piplus)
F_V^\pi(s_{\pi\pi})\ . 
\en 
with $s_{\pi\pi}=(p_\pizero+p_\piplus)^2$.
One can then write the unitarity relation in the form
\bea\lbl{discgammapipi}
&& \disc[\fzero]_{\gamma\pi\pi}=
\theta(s-4\mpid){-e\over\Delta_{\eta\pi}}\times
      {1\over2}\sum_{\lambda=\pm1}\nonumber\\
&&  \phantom{\disc[\fzero]_{\gamma\pi\pi}}
\int d\hbox{Lips}_3\,F_V^\pi(s_{\pi\pi}) \,e_\gamma(\lambda)\cdot (p_\pizero-p_\piplus)
T^*_{\eta\pi^+\to \gamma(\lambda)\pi^0\pi^+}\ .
\ena
\end{itemize}
(where $d\hbox{Lips}_3$ is the three-body Lorentz invariant
phase-space measure).
A resonant contribution from the $a_0(980)$ to the amplitude
$\eta\pi\to\gamma\pi\pi$ which appears in eq.~\rf{discgammapipi} is
possible. However, a suppression of this contribution in the region
below 1 GeV is expected because of the three-body phase space.

In summary, below the $K\Kbar$ threshold, the dominant contribution
is from $n=\pi^0\pi^+$ (enhanced by the $\rho(770)$ resonance) for the
vector form factor and from $n=\eta\pi^+$ (enhanced by the $a_0(980)$)
for the scalar form factor. We will use this result in the sequel in
order to evaluate the two form factors with the help of dispersion
relations. In order to suppress the sensitivity of the integrals to
the region $s \ge 1$ $\hbox{GeV}^2$ it is necessary to introduce
subtractions. We now recall the results which have been obtained in
ChPT at NLO from which we will be able to estimate the subtraction
constants. 

\section{Results from ChPT at order $p^4$}\lblsec{resultschpt}
The $\eta\pi$ form factors have been computed in ChPT at next-to-leading 
order in refs.~\cite{Neufeld:1994eg,Scora:1995sj}. We collect
below some of the results which are relevant to our study.

\subsection{Form factors at $s=0$}
Consider first the form factors at $s=0$. At leading order in ChPT
they are simply equal to the $\pi^0-\eta$ mixing angle,
\be\lbl{fplus0LO}
\left.f_+^{\eta\pi}(0)=f_0^{\eta\pi}(0)\right\vert_{LO}=\epsilon
\en
where $\epsilon$ is given in~\rf{epsilon}.
At the same order, $\epsilon$ can be determined using the experimental
values of the pseudoscalar meson masses $m_\piplus$, $m_\pizero$,
$m_\kplus$, $m_\kzero$, together with Dashen's low-energy
theorem~\cite{Dashen:1969eg}, 
\be\lbl{dashenth}
\left.\mkplusd- \mkzerod\right\vert_{EM} = m^2_\piplus-m^2_\pizero
+O(e^2p^2),\ O(p^4)
\en
which gives
\be\lbl{epsLOvalue}
\left.\epsilon\right\vert_{LO}\simeq 0.99\cdot10^{-2}\ .
\en
The corrections of order $p^4$, including also the electromagnetic
$e^2p^2$  piece, were written in ref.~\cite{Neufeld:1994eg} (see
also~\cite{Cirigliano:2001mk}) in the following form 
\bea\lbl{fplus0NLO2}
&& \left.f_+^{\eta\pi}(0)=f_0^{\eta\pi}(0)\right\vert_{LO+NLO}=
\epsilon- {2\epsilon\over3\Delta_{\eta\pi}\fpid}\bigg[
64 \Delta_{K\pi}^2(3L_7^r+L_8^r)
\nonumber\\
&&\quad
 -\metad\, \Delta_{K\pi} L_\eta-2\mkd(\mkd-2\mpid)L_K+\mpid(\mkd-3\mpid)L_\pi
-{2\mkd \Delta_{K\pi} \over16\pi^2}\bigg]
\nonumber\\
&&\quad
+{2\sqrt3 e^2\mkd\over27\Delta_{\eta\pi}} 
\bigg[{2}(2S_2^r+3S_3^r)-9Z(L_K+{1\over16\pi^2})\bigg]
\ena
with
\be
L_P\equiv {1\over16\pi^2}\log{\mpd\over\mud}\ ,
\en 
(an additional small electromagnetic contribution, proportional to
$\mpid/\Delta_{\eta\pi}$ has been neglected).
Here, $L_7^r$, $L_8^r$ are the standard  low-energy coupling
constants of the strong chiral Lagrangian at $O(p^4)$~\cite{gl85},
while $Z$, $S_2^r$, $S_3^r$ are $O(e^2)$ and $O(e^2p^2)$ electromagnetic
couplings~\cite{Urech:1994hd,Neufeld:1995mu}. One can also express
$\fplus(0)$ in terms of the two $\pi^0-\eta$ mixing angles
$\epsilon_1$, $\epsilon_2$ introduced in ref.~\cite{gl85}, 
\be
\left.f_+^{\eta\pi}(0)=f_0^{\eta\pi}(0)\right\vert_{LO+NLO}={1\over2}(\epsilon_1+
\epsilon_2)
\en
(the electromagnetic contributions to $\epsilon_1$, $\epsilon_2$ can be found
in~\cite{Ananthanarayan:2002kj}). 
In using eq.~\rf{fplus0NLO2}, one must also update the determination of
the mixing angle $\epsilon$, e.g. from the pseudoscalar meson masses,
including $O(p^4)$ and $O(e^2p^2)$ contributions. At present,
however, the $O(e^2p^2)$ LEC's which play an important numerical role
for the meson masses are not known in a model independent way.
Fortunately, it was observed in ref.~\cite{Neufeld:1994eg}
that a very simple relation holds between $\fplus(0)$ and 
an isospin-breaking difference involving the $K^0\pi^\pm$ and $K^\pm\pi^0$  
vector form factors. In its updated form, it can be written in terms of
the form factors $\tilde{f}_+^{K\pi}$  defined in ref.~\cite{Cirigliano:2001mk} 
to include the isospin-breaking effects from QCD and part of the radiative
corrections as
\be\lbl{NR}
f_+^{\eta\pi}(0)={1\over\sqrt3}\delta^{K\pi} +O(p^6),\quad 
\delta^{K\pi}=\frac{\tilde{f}_+^{\kplus\pizero}(0)}{
  \tilde{f}^{\kzero\piplus}_+(0)}-1\ . 
\en 
This relation was not used in earlier work on the second-class
$\eta\pi$ amplitudes because the precision of the experimental results
on the $K_{l3}$ form factors was insufficient. The situation has
considerably improved in recent years and, using the  
averaged experimental result quoted in the review~\cite{Antonelli10}, 
\be
\left.\delta^{K\pi}\right\vert_{exp}=0.027\pm0.004
\en 
one obtains
\be\lbl{f+(0)exp}
\left.f_+^{\eta\pi}(0)\right\vert_{NLO+exp}=(1.56\pm0.23)\,10^{-2}\ .
\en
Clearly, this is a very significant enhancement as compared to the
leading order result. 
%modif
For comparison, the following range of values was quoted in
ref.~\cite{Neufeld:1994eg}: $f_+^{\eta\pi}(0)=[1.22-1.37]\cdot
10^{-2}$ based on the chiral expression~\rf{fplus0NLO2}, using the LO
result for $\epsilon$ and order of magnitude estimates for the
electromagnetic coupling constants. A result completely compatible
with~\rf{f+(0)exp} can be obtained if the value of the quark mass
ratio $\epsilon$ is enhanced from its LO result by 20-30\% due to NLO
effects. There are indications that this could indeed be the case from
model estimates as well as lattice QCD calculations of the chiral
corrections to Dashen's low-energy theorem (see e.g. the FLAG
review~\cite{Aoki:2013ldr}). We will return to the question of the
quark mass ratio in sec.~\sect{5.4} in connection with $\eta\to 3\pi$
decay. The fact that compatible results are obtained using either the
low-energy theorem relation~\rf{NR} or the chiral
expression~\rf{fplus0NLO2}, which have different $O(p^6)$ corrections,
is an indication that $O(p^6)$ corrections should be of natural size
(5-10\%, say) in spite of the large size of the NLO correction.

\subsection{Vector form factor}
We reproduce below the expression of  the vector form factor
$\fplus(s)$ from ref.~\cite{Neufeld:1994eg}, in a slightly
re-expressed form, which involves the scalar loop functions 
$\bar{J}_{PQ}(s)$~\cite{gl85} 
\be
\bar{J}_{PQ}(s)={s\over16\pi^2}\int_{(m_P+m_Q)^2}^\infty ds'
    {\sqrt{\lambda_{PQ}(s')}\over (s')^2 (s'-s)}\ .
\en
The result for $\fplus$ reads
\bea\lbl{fplusp4}
&& \fplus(s)= \fplus(0)
 +{\epsilon\over12\fpid}\bigg\{ (s-4\mkd)\jbar_{KK}(s)
+2(s-4\mpid)\jbar_{\pi\pi}(s)\nonumber\\
&& \phantom{H_1(s)= -{1\over12\fpid}}
+s\left[24L_9^r-L_K-2L_\pi-{1\over16\pi^2}\right] \bigg\}\ .
\ena
Eq.~\rf{fplusp4} allows one to deduce the value of the
derivative of the vector form factor at $s=0$ which will serve us,
together with $f_+^{\eta\pi}(0)$, to normalise  the dispersive
construction of the form factor. One finds,
\be\lbl{dotf+(0)exp}
 \dot{f}_+^{\eta\pi}(0)= {\epsilon \over12\fpid}
\left(24L_9^r(\mu)-L_K-2L_\pi-{3\over16\pi^2}\right)
=\epsilon\, (1.70\pm0.10)\ \hbox{GeV}^{-2}.
\en
In the last equality, we have used the value of the coupling $ L_9^r$
given by~\cite{Bijnens:2002hp}, $L_9^r(m_\rho)=(5.93\pm0.43)\,10^{-3}$.

\subsection{Scalar form factor}\lblsec{scalarffp4}
Next, the scalar form factor $\fzero$ from ref.~\cite{Neufeld:1994eg}
can be expressed as follows:  
\be\lbl{fzerop4}
\fzero(s)= f_+^{\eta\pi}(0) + {\epsilon\,\over\fpid}H_\epsilon(s)
+ {e^2\over\Delta_{\eta\pi}} H_e(s)
\en
with
\bea\lbl{fzerop4b}
&&  H_\epsilon(s)= {\mpid\over3} \jbar_{\eta\pi}(s) 
+{1\over4}\bigg\{
(3s-4\mkd)\jbar_{KK}(s)
+s\left[16L_5^r-3L_K-{3\over16\pi^2}\right]\bigg\}\nonumber\\ 
&&  H_e(s)={\sqrt3\over18}\bigg\{ -3Z(3s-4\mkd)\jbar_{KK}(s)\nonumber\\
&&\phantom{s H_3(s)=} 
+s\Big[ -2(2S_2^r+3S_3^r)+9Z\Big(L_K+{1\over16\pi^2}\Big)\Big] \bigg\}
\ena
One expects that the scalar form factor, evaluated at the point
$s=\Delta_{\eta\pi}=\metad-\mpid$ should satisfy a Callan-Treiman
relation~\cite{Callan:1966hu,Dashen:1969bh} 
\be\lbl{CTrel}
\fzero(\Delta_{\eta\pi})=-{F^{(3)}_\eta\over F^{(3)}_\pizero} + \Delta_{CT},\ 
\Delta_{CT}=O(\mpid)\ , 
\en
where $F_P^{(3)}$ is defined from the matrix element of the axial
current $j_{\mu,5}^3$
\be
\braque{0\vert j_{\mu,5}^3(0)\vert P}=i p_\mu F_P^{(3)}
,\ P=\eta, \pi^0\ .
\en
Indeed, using the  expressions for $F^{(3)}_\eta$, $F^{(3)}_\pizero$ computed
in ref.~\cite{Neufeld:1995mu} at chiral order $p^4$ and $e^2 p^2$
one obtains that $\Delta_{CT}$ is proportional to $\mpid$ and reads
\be\lbl{ctdeviation}
\Delta_{CT}= 
\epsilon\,{\mpid\over3\fpid}\left[
\jbar_{\eta\pi}(\Delta_{\eta\pi})-3\jbar_{KK}(\Delta_{\eta\pi})\right]
+e^2{2Z\mpid\over\sqrt3\Delta_{\eta\pi}} \jbar_{KK}(\Delta_{\eta\pi})\ .
\en
The Callan-Treiman relation~\rf{CTrel} could be used to evaluate
$\fzero(\Delta_{\eta\pi})$ rather precisely if the ratio ${F^{(3)}_\eta/
  F_\pi}$ were known accurately from lattice QCD.
For now, we must rely only on the chiral expansion and, as in the case
of the vector form factor, we will use the value of the derivative of
the scalar form factor at $s=0$ as an input to the dispersive
calculation. Using the chiral expression~\rf{fzerop4b} with the value
of the LEC $L_5$: $L_5^r(m_\rho)=(1.20\pm0.05)\cdot10^{-3}$ deduced
from the ratio ${F_K/F_\pi}=1.192\pm0.005$~\cite{Aoki:2013ldr} one
obtains
\be\lbl{dotfzero}
\dotfzero(0)=  ((0.404\pm 0.025)\,\epsilon-5.49\cdot10^{-4})\quad
\hbox{GeV}^{-2} 
\en
where the last term is the electromagnetic contribution evaluated
%modif
using  resonance modelling estimates~\cite{Ananthanarayan:2004qk} of
the couplings $S_2$, $S_3$.

As a final remark, we note that in ChPT at NLO the discontinuities
of the form factors (which coincide with the imaginary parts at this
order) are generated by the functions $\jbar_{\pi\pi}$,
$\jbar_{\eta\pi}$ and $\jbar_{KK}$. As a simple check, we show in
appendix~\sect{verifdiscp4}  that one recovers these NLO 
results  from the general unitarity relations as given in
sec.~\sect{defsbasic} using the chiral $O(p^2)$ expressions for the
form factors and the four-meson amplitudes which enter in these
relations.   

\section{Absence of anomalous thresholds in $\eta\pi$ form factors in
  a toy model}\lblsec{remarkanomth}
%%%%%%%%%%%%%%%%%%
\begin{figure}[hb]
\bc
\includegraphics[width=0.5\linewidth]{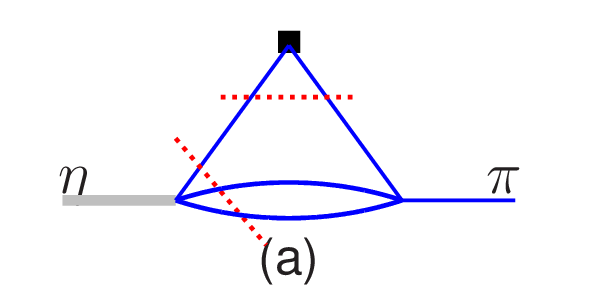}\includegraphics[width=0.5\linewidth]{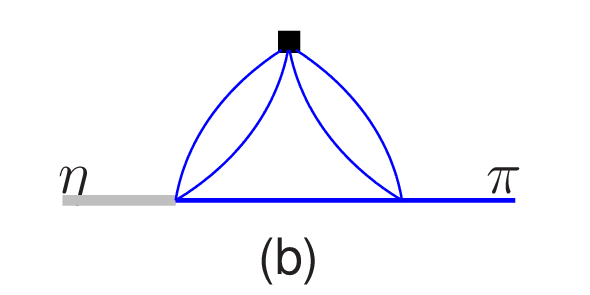}
\caption{\small Two types of  diagrams contributing to $\eta\pi$ form factors
and involving one $\eta\to3\pi$ vertex.}
\label{fig:vertextypes}
\ec
\end{figure}
%%%%%%%%%%%%%%%%%%
A usually accepted property  of form factors involving two stable 
particles (like the pion or the nucleon electromagnetic form factors)
is that they can be defined as  analytic functions of the energy
variable $s=(p_1-p_2)^2$, with a cut along the positive real axis, the
discontinuity along this cut being given by unitarity relations
(e.g.~\cite{Barton:1965}).  

\begin{figure}[ht]
\bc
\includegraphics[width=0.5\linewidth]{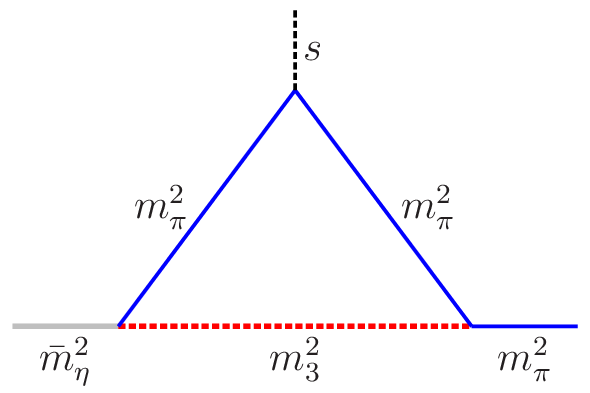}
\caption{\small Triangle diagram in eq.~\rf{triangleint}.}
\label{fig:triangle}
\ec
\end{figure}

Here, we wish to consider the $\eta\pi$ form factors $\fplus$,
$\fzero$ and, since the $\eta$ meson is unstable, one should be concerned
about the presence of anomalous thresholds. We illustrate in
Fig.~\fig{vertextypes} the two types of diagrams which involve the
$\eta\to3\pi$ decay amplitude at one vertex. We will consider here
contributions of the form of Fig.~\fig{vertextypes}(a) in which the
normal threshold is $s_{th}=4\mpid$. The contributions of the second
type, as shown in Fig.~\fig{vertextypes}(b) have a much higher
normal threshold $s_{th}=16\mpid$ and the discontinuity function is
expected to be very much suppressed in the region $s\lapprox 1$
$\hbox{GeV}^2$ because of the four-body phase space.
Let us discuss here the question of the anomalous threshold in the toy
model case where Fig.~\fig{vertextypes}(a) represents a  Feynman diagram
with local vertices. The form factor can then be represented as an
integral (e.g.~\cite{Gasser:1998qt})
\be\lbl{triangleint}
f^{\eta\pi}(s)=\int_{4\mpid}^\infty dt'\,\rho(t') K^{\eta\pi}(t',s)+\cdots
\en
where $K^{\eta\pi}(t',s)$ corresponds to the simple triangle Feynman
diagram with external momenta $p_1^2=\bar{m}^2_\eta$, $p_2^2=\mpid$, 
$p_3^2=s$ and internal masses $m_1^2=m_2^2=\mpid$, $m_3^2\equiv t'$
(see fig.~\fig{triangle}), and the weight function is 
\be
\rho(t')={1\over16\pi^2}\sqrt{1-{4\mpid\over t'}}\ .
\en
%%%%%%%%%%%%%%%%%%
%%%%%%%%%%%%%%%%%%
In eq.~\rf{triangleint}, additional terms (including the UV divergent
ones ) have been omitted since they are not concerned with the
possibility of an anomalous threshold. We will vary the value of
$\bar{m}^2_\eta$ and, since it can be considered as an energy
variable, the amplitude may be defined by appending an infinitesimal
positive imaginary part to it, 
i.e. 
\be
\bar{m}^2_\eta\equiv \lim_{\epsilon\to0}(\bar{m}^2_\eta+i\epsilon)\ .
\en 
We first take $\bar{m}^2_\eta$ to be sufficiently
small (e.g. $\bar{m}^2_\eta=\mpid$) such that an ordinary dispersion
relation (DR) holds and then increase $\bar{m}^2_\eta$ until it reaches the
physical value $\metad$. The ordinary DR for the triangle graph reads,
\be\lbl{ODR}
K^{\eta\pi}(t',s)={1\over\pi}\int_{4\mpid}^\infty ds'\,
  {L^{\eta\pi}(t',s')\over s'-s}
\en
where $L^{\eta\pi}(t',s')$ is the  discontinuity function of the triangle graph,
\bea\lbl{Detapi}
&& L^{\eta\pi}(t',s')={1\over16\pi\lambda^{1\over2}(s',\bar{m}^2_\eta,\mpid)}
\log{a+b\over a-b},\nonumber\\
&& a=s'-(\bar{m}^2_\eta+3\mpid-2t'),\
b=\sqrt{1-{4\mpid\over
    s'}}\lambda^{1\over2}(s',\bar{m}^2_\eta,\mpid)\ .
\ena
We note that the discontinuity of the form factor is then given as an
integral over $L^{\eta\pi}(t',s)$ 
\be\lbl{discffint}
\disc[f^{\eta\pi}(s)]=\int_{4\mpid}^\infty dt' \rho(t')
L^{\eta\pi}(t',s)+\cdots 
\en
As discussed in refs.~\cite{Mandelstam:1960zz,Oehme:1961} the presence
of anomalous thresholds can be inferred from studying the motion of
the singularities of the function $L^{\eta\pi}$ upon
varying $\bar{m}^2_\eta$: if one of the singularities crosses the
unitarity cut, it is then necessary to deform the contour in the
dispersion representation~\rf{ODR}, in order to properly define its
analytical continuation as a function of $\bar{m}^2_\eta$, thereby
introducing an anomalous threshold. In the present case~\rf{Detapi},
the singularities of the function $L^{\eta\pi}(t',s)$ are given by the
solutions of the  equation $a^2-b^2=0$, which is quadratic in $s$
\be\lbl{singuleq}
t' s^2 + t'(t'-3\mpid-\bar{m}^2_\eta-i\epsilon)s 
+\mpid(\mpid-\bar{m}^2_\eta-i\epsilon)^2=0\ .
\en 
%%%%%%%%%%%%%%%%%%
\begin{figure}[ht]
\bc
\includegraphics[width=\figwidth]{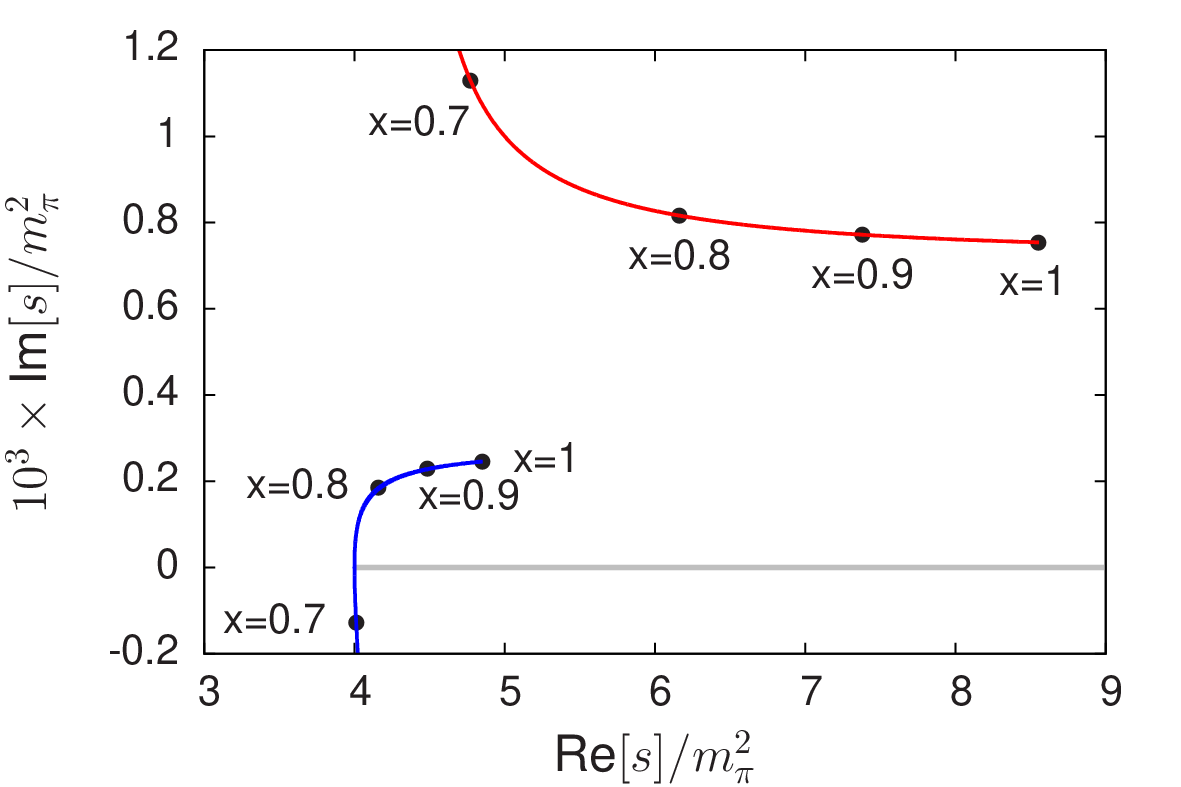}
\caption{\small Illustration of the motion of the two singularities of
  the discontinuity function $L^{\eta\pi}(t',s)$ in the $s$ plane for a
  given value of $t'$ when varying the external mass squared
  $\bar{m}^2_\eta\equiv x\metad$. The plot corresponds to
  $t'=5\mpid$, $\epsilon=10^{-3}\mpid$. }
\label{fig:anomth1}
\ec
\end{figure}
%%%%%%%%%%%%%%%%%%
Let us consider three cases, depending on the value of the mass
variable $t'$
\begin{enumerate}
\item $t'=4\mpid$: In this case, the two singularities coincide and
  are given by
\be
s^\pm= {1\over2}( \barmetad-\mpid +i\epsilon),
\en
which is above the unitarity cut.
\item $4\mpid < t' \le {1\over2}(\barmetad-\mpid)$: This is the most
  interesting situation, the motion of the
  two singularities $s^\pm(\bar{m}^2_\eta)$ as a function of
  $x=\bar{m}^2_\eta/\metad$ is illustrated in
  Fig.~\fig{anomth1}. The figure shows that while 
  $s^+(\bar{m}^2_\eta)$ remains above the real axis, the other
  singularity does cross the real axis very close to $4\mpid$. It is easy
  to see that the crossing occurs when $\bar{m}^2_\eta=\mpid+2 t'$
  and the value of the crossing point is
\be
s^-(\mpid+2 t')=4\mpid\left(1-{\epsilon^2\over
  4t'(t'-4\mpid)}
\right)\ ,
\en
which is located strictly {\em below} the normal threshold.  
\item ${1\over2}(\barmetad-\mpid) < t' \le (\barmeta-\mpi)^2$: In this
  case $s^+( \barmetad)$ remains above the real axis and  $s^-(
  \bar{m}^2_\eta)$ remains below it.
\end{enumerate}
For larger values of $t'$, it is easily verified that $s^\pm(\barmetad)$
do not come close to the unitarity cut. The conclusion of this
discussion is that, for a Feynman diagram,  the amplitude as given
from Fig.~\fig{vertextypes}(a) does not involve any anomalous
threshold. We will argue in the next section that the same conclusion
holds in a more realistic approach where the $\eta\to 3\pi$ amplitude
is given from Khuri-Treiman equations solutions.
The fact that the $\eta$ meson is unstable, i.e. $\meta
>3 \mpi$, manifests itself in the violation of real analyticity,
the discontinuity function is complex and does not coincide with the
imaginary part of the diagram (the latter, indeed, is given by the sum
of two Cutkosky contributions, corresponding to two physically allowed
ways of cutting the diagram; see Fig.~\fig{vertextypes}(a)). 

\section{Dispersive evaluation of  $\fplus$ }
We have argued in sec.~\sect{defsbasic} that the dominant contribution
to the discontinuity of $\fplus(s)$, with $s\lapprox 1$ \Gevd, was from
the $n=\pi^0\pi^+$ state. Then, $\disc[\fplus]$ was found to be
proportional to the pion vector form factor $F_V^\pi$, which is well
known from experiment, and to the $l=1$ projection of the
$\eta\pi^+\to\pi^0\pi^+$ amplitude. In order to evaluate this
amplitude, we will make use of the work of
refs.~\cite{Kambor:1995yc,Anisovich:1996tx}, who developed and solved
a set of Khuri-Treiman~\cite{Khuri:1960zz} equations and
applied the results in the physical region of the $\eta\to3\pi$
decay. We briefly recall the main features of this formalism below (a
very detailed account can be found in the thesis~\cite{Lanz:2011}). As it
encodes the analyticity properties of the $\eta\pi^+\to\pi^0\pi^+$
amplitude, this formalism should be suitable for evaluating the
amplitude in the partly unphysical region needed for computing
$\disc[\fplus]$. 

\subsection{Brief review of the Khuri-Treiman formalism}\lblsec{KTreview}
The Khuri-Treiman (KT) equations implement dispersion relations,
crossing symmetry and unitarity in the
approximation where a single state, $n=\pi\pi$, is retained in the
unitarity relations. This approximation is acceptable when the
Mandelstam variables $s$, $t$, $u$ are smaller than 1 ${\rm GeV}^2$ in
magnitude. It was noted in refs.~\cite{Kambor:1995yc,Anisovich:1996tx}
that, in the same region, the contributions from the discontinuities
of the partial-waves $J\ge2$ can also be neglected in the dispersion
relations such that the decomposition theorem~\cite{Stern:1993rg} may
be applied to the $\eta\to 3\pi$ amplitude. As a result, it can be
expressed in terms of three functions  of a single variable, $M_I(w)$, 
\bea\lbl{eta3pidecomp}
&& T_{\eta\pi^0\to\pi^-\pi^+}(s,t,u)= T_{\eta\pi^+\to\pi^0\pi^+}(t,s,u)= 
 - \epsilon_L\times\!\Big[M_0(s)+(s-u)M_1(t) \nonumber\\
&&\phantom{T_{\eta\pi^0\to\pi^-\pi^+}(s,t,u)=} 
+(s-t)M_1(u)+M_2(t)+M_2(u)-{2\over3}M_2(s) 
\Big]
\ena
with
\be
\epsilon_L={1\over Q^2}{\mkd\over\mpid}\,{\mkd-\mpid\over3\sqrt3\fpid},\quad
Q^2={m_s^2-\hat{m}^2\over m_d^2-m_u^2}\ .
\en
Based on usual Regge phenomenology for estimating the asymptotic behaviour, 
it was concluded in ref.~\cite{Anisovich:1996tx} that $M_0$, $M_2$
should obey converging DR's with two subtractions and $M_1$ a
converging DR with a single subtraction,
\bea\lbl{disprelMI}
&& M_0(w)=\tilde\alpha_0 + \tilde\beta_0 w
+{w^2\over\pi}\int_{4\mpid}^\infty ds'\,
\frac{\disc[M_0(s')]}
{(s')^2(s'-w)}         \nonumber\\
&& M_1(w)= {w\over\pi} \int_{4\mpid}^\infty ds'\,
\frac{\disc[M_1(s')]}
{s' (s'-w)}            \nonumber\\
&& M_2(w)= {w^2\over\pi}\int_{4\mpid}^\infty ds'\,
\frac{\disc[M_2(s')]}
{(s')^2(s'-w)}\ . 
\ena 
In writing these DR's one has further made use of  freedom 
to redefine $M_I$ by linear functions without modifying the
$\eta\pi\to\pi\pi$ amplitude in eq.~\rf{eta3pidecomp} because of the
constraint $s+t+u=\metad+3\mpid \equiv 3s_0$. 
The functions $M_I(w)$ are analytic in the complex $w$ plane except for a
cut along the real axis along $[4\mpid,\infty]$. The discontinuity
along this cut is obtained from the unitarity relations of the $J=0,1$
partial-wave projections of the $\eta\pi\to\pi\pi$ amplitudes and they
read,
\be\lbl{discMI}
\disc[M_I(s)]_{\pi\pi} =\theta(s-4\mpid)\,
e^{-i\delta_I(s)}\sin\delta_I(s) (M_I(s+i\epsilon)+\hat{M}_I(s))\ .
\en
In eq.~\rf{discMI} the functions $\hat{M}_I$ are linear combinations
of the angular integrals,
\be\lbl{znMIdef}
\braque{z^n M_I}(s)={1\over2}\int_{-1}^1 dz z^n M_I(t(s,z))
\en
with
\be
t(s,z)={1\over2}(3s_0-s+\kappa(s)z),\
\kappa(s)=\sqrt{1-4\mpid/s}\lambetapi{s}\ .
\en
The explicit expressions of $\hat{M_I}$ in terms of the angular
integrals read~\cite{Anisovich:1996tx}
\bea\lbl{defMIhat}
&&\hat{M}_0={2\over3}\braque{M_0}+{20\over9}\braque{M_2}+2(s-s_0)\braque{M_1}
+{2\over3}\kappa \braque{z M_1}\nonumber\\
&&\hat{M}_1=\kappa^{-1}\left\{ 3\braque{z M_0}-5 \braque{zM_2}
+   {9\over2}(s-s_0)\braque{z M_1}
+   {3\over2}\kappa \braque{z^2 M_1}\right\}\nonumber\\
&& \hat{M}_2= \braque{M_0}+{1\over3}\braque{M_2}-{3\over2}(s-s_0)\braque{M_1}
-{1\over2}\kappa \braque{z M_1}\ .
\ena
Eqs.~\rf{disprelMI} and~\rf{discMI}~\rf{defMIhat} are a first form of
the Khuri-Treiman integral equations for the $\eta\pi\to\pi\pi$ amplitude.
\subsection{Singularities of the functions $\hat{M}_I$}
\begin{figure}[ht]
\bc
\includegraphics[width=\figwidth]{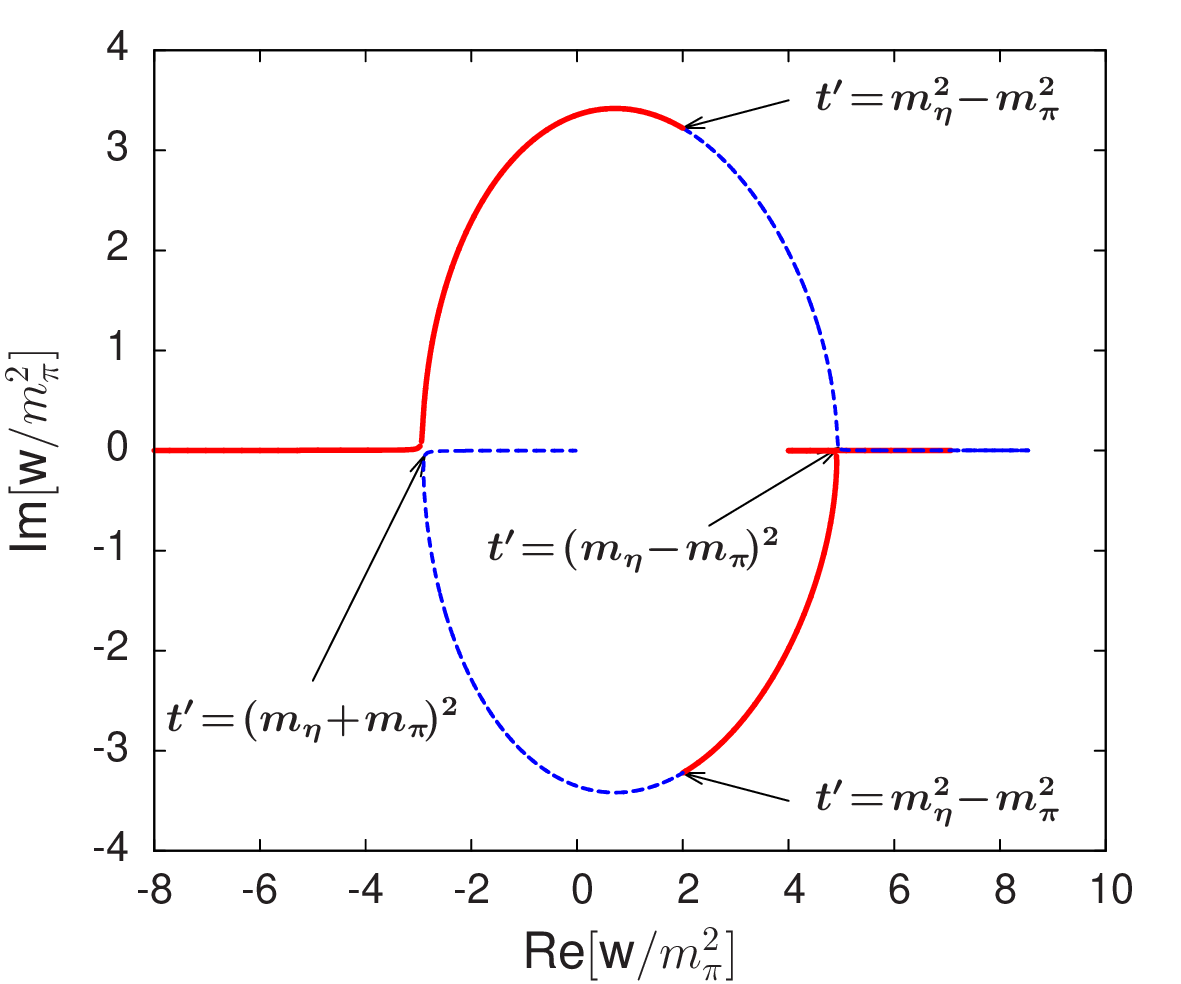}
\includegraphics[width=\figwidth]{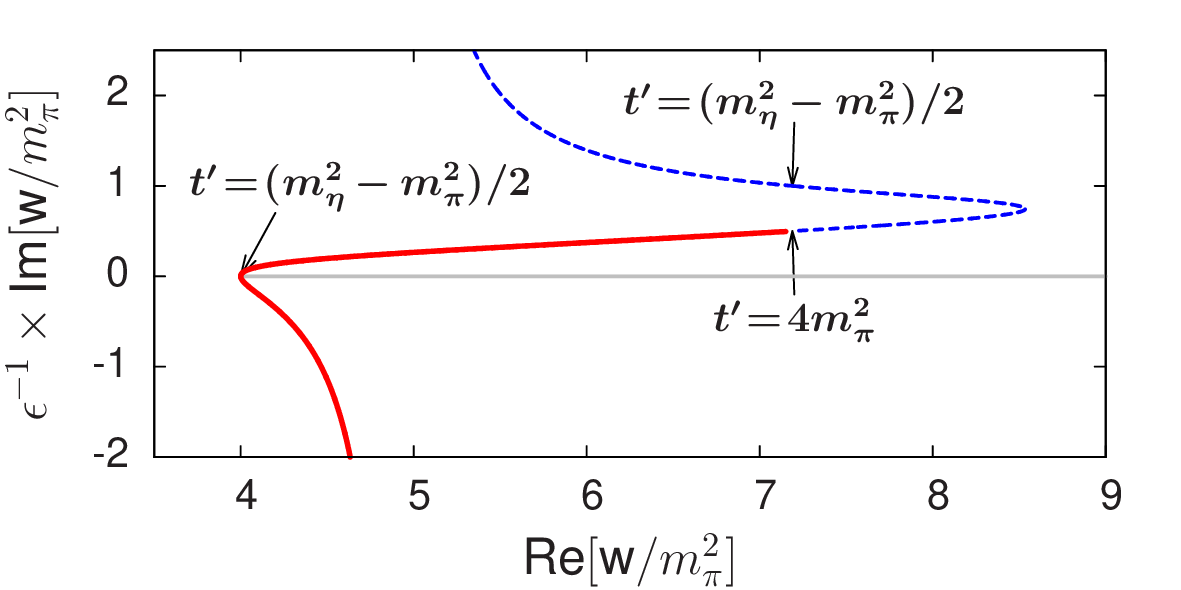}
\caption{\small  Complex cut ${\cal C}$ of the functions
  $\hat{M}_I(w)$ (see eq.~\rf{cutmhat}: the solid (dashed) curves
  correspond to the points which satisfy $w=t^-(t')$ ($t^+(t')$).  
  The lower figure shows an enlarged view of the vicinity of the
  unitarity cut.}  
\label{fig:cxcut}
\ec
\end{figure}

\lblsec{singmhat}
Using the representation of $T_{\eta\pi\to\pi\pi}$ based on the
decomposition theorem~\rf{eta3pidecomp}, we can now write the
discontinuity of the form factor $\fplus$ as
\be\lbl{discfplusM1}
\disc[\fplus(s)]_{\pi\pi}=\theta(s-4\mpid){(s-4\mpid)^{3\over2}\over 48\pi\sqrt{s}}
 \epsilon_L F_V^\pi(s-i\epsilon)( M_1(s+i\epsilon)+\hat{M}_1(s))\ .
\en
For completeness, let us mention the analogous relation for the scalar
form factor,
\be
\disc[\fzero]_{\pi\pi}=\theta(s-4\mpid){(s-4\mpid)^{1\over2}\over 16\pi\sqrt{s}}
{\Delta_{\pizero\piplus}\over\Delta_{\eta\piplus}} \epsilon_L f_0^{\pi\pi}(s-i\epsilon)
(M_2(s+i\epsilon)+\hat{M}_2(s))\ .
\en
Adapting the discussion of sec.~\sect{remarkanomth} about the presence
of anomalous thresholds to the present, more realistic situation,
requires one to investigate the singularities of the functions
$\hat{M}_I(w)$ i.e. of the angular integrals given in~\rf{znMIdef}. 
This can be done by inserting the dispersive representations of the
functions $M_I$ (eqs.~\rf{disprelMI}) into the angular integrals
(eqs.~\rf{znMIdef}) from which  one obtains an expression of the functions
$\braque{z^nM_I}$ as integrals over kernels $K^{(n)}(t',w)$, $P^{(n)}(t',w)$
\bea\lbl{kernelsproj}
&& \braque{M_0}(w)=\tilde\alpha_0 +{1\over2}(3s_0-s)\tilde\beta_0
-{1\over\pi}\int_{4\mpid}^\infty dt'
K^{(0)}(t',w)\,\disc[M_0(t')]\nonumber\\
&& \braque{zM_0}(w)={1\over6}\kappa(w)\tilde\beta_0
-{1\over\pi}\int_{4\mpid}^\infty dt'
K^{(1)}(t',w)\,\disc[M_0(t')]\nonumber\\
&& \braque{z^n M_2}(w)= -{1\over\pi}\int_{4\mpid}^\infty dt'
K^{(n)}(t',w)\,\disc[M_2(t')]\nonumber\\
&& \braque{z^n M_1}(w)= -{1\over\pi}\int_{4\mpid}^\infty dt'
P^{(n)}(t',w)\,\disc[M_1(t')]\ .
\ena
The kernels which are needed here are given explicitly in
appendix~\sect{kernels}. They involve the logarithmic function
\be\lbl{mainlog}
L(t',w)={1\over2}\int_{-1}^1 dz\,{1\over t(w,z)-t'}\ .
\en
which controls their singularity structure.
When performing the angular integration in
eqs.~\rf{znMIdef},~\rf{mainlog} 
one must keep in mind that the path of integration from $z=-1$ to
$z=1$ must eventually be deformed in order not to intersect with the
cut of the functions $M_I$, i.e. $4\mpid \le t(w,z)< \infty$, as is
explained in detail in ref.~\cite{Kambor:1995yc}. The following 
expression of the logarithmic function exactly encodes these
prescriptions on the integration path in a simple way,
\be\lbl{basiclog}
L(t',w)=   
{\log(t'-t^+(w))-\log(t'-t^-(w))\over t^+(w)-t^-(w)}
\en
where
\be\lbl{tplusminu}
t^\pm(w)={1\over2}\Big(\metad+i\epsilon+3\mpid-w \pm
\sqrt{1-{4\mpid\over s}}\lambda^{1\over2}(\metad+i\epsilon,\mpid,w)\Big)
\en
displaying explicitly the $i\epsilon$ prescription.
From the form of the logarithmic
function~\rf{basiclog},~\rf{tplusminu} we can infer the following
consequences: 
\begin{enumerate}
\item Absence of anomalous thresholds: Inserting the representation of 
$\hat{M}_1$, $\hat{M}_0$ in terms of the kernels~\rf{kernelsproj} one
  obtains an integral representation of the form factor
  discontinuities $\disc[\fplus]$, $\disc[\fzero]$ in terms of the
  logarithm function~\rf{basiclog} which is analogous to
  eq.~\rf{discffint}. Furthermore, the singularities of the
  logarithms are exactly the same. Therefore, the conclusion about
  the absence of anomalous thresholds, as discussed in
  sec.~\sect{remarkanomth}, applies also in the present realistic
  situation.
\item Cuts ${\cal C}$ of the $\hat{M}_I$ functions: They are given,
  using the integral representations in terms of kernels, as the
  ensemble of the singularities of the logarithmic function~\rf{basiclog}
  (see~\cite{Kennedy1961}) i.e. the points $w$ which satisfy
  $t^\pm(w)=t'$. This relation can be recast as a quadratic equation 
  in $w$ identical to~\rf{singuleq}. Therefore,
\be\lbl{cutmhat}
{\cal C}=\{w: w=t^\pm(t'),\ 4\mpid\le t' < \infty\}, 
\en 
The curve~\rf{cutmhat} includes the negative real axis, it extends
into the complex plane and 
approaches infinitesimally close to
  the unitarity cut on the positive real axis in the region 
  $4\mpid\le w \le (\meta-\mpi)^2$ without, however,
  crossing it: this is shown on fig.~\fig{cxcut}. 
As a consequence, the functions $\hat{M}_I(s')$ are unambiguously
defined on the real axis in the integration range $4\mpid\le s'
<\infty$.   
\end{enumerate}
%modif
%%%%%%%%%%%%%%%%%%
The point $w=\mmd \equiv (\meta-\mpi)^2$ requires  special
attention. At this point the difference between $t^+(w)$
and $t^-(w)$ is infinitesimal but their imaginary parts have different
signs 
\be
t^\pm(\mmd)=\mpi(\meta+\mpi)
         \pm{i\sqrt\epsilon\over2}\sqrt{\mpi(\meta-\mpi)}
\en
which implies that the logarithmic function $L(t',w)$ diverges 
when $\epsilon$ goes to zero  and $t'$ lies in the range $4\mpid\le t'
< \mpi(\meta+\mpi)$  
\be
\lim_{w\to\mmd} L(t',w)={2\pi\over \sqrt{\epsilon}}{\theta(\mpi(\meta+\mpi)-t')
\over\sqrt{\mpi(\meta-\mpi)}}
\en
which induces a divergence in the functions $\hat{M_I}(w)$ when $w\to\mmd$
\be
\lim_{w\to\mmd} \hat{M}_0(w),\, \hat{M}_2(w)\sim O(\epsilon^{-{1\over2}}),\quad
\lim_{w\to\mmd} \hat{M}_1(w)\sim O(\epsilon^{-{3\over2}})\ .
\en
However, integrals over $\hat{M}_I$ (as  in
eqs.~\rf{kernelsproj},~\rf{Ihat}) remain finite in the $\epsilon\to0$
limit~\cite{Kambor:1995yc} such that the functions $M_I(w)$ themselves
do no exhibit any divergence. 
The numerical treatment of the integrations involving $\hat{M}_I$ is
the delicate part of solving the KT equations. When the integration
variable is close to $\mmd$ one must perform expansions in powers of
$\sqrt{\mmd-t'}$ and use the analytical expressions for the integrals
of the functions $\log(t'-t^\pm(w))/(\mmd-t'+i\epsilon)^{n+1/2}$
(appearing in eq.~\rf{kernelsproj}) and 
$1/((t'-w-i\epsilon)(\mmd-t'+i\epsilon)^{n+1/2})$ (in eq.~\rf{Ihat}).
Analogous singular integrations  appear in the dispersive representation of the
vector form factor $\fplus$ as its discontinuity involves $\hat{M}_1$.

\subsection{Matching with ChPT }
In the form given by eqs.~\rf{disprelMI}~\rf{discMI}~\rf{defMIhat}
the KT equations are linear integral equations with a singular Cauchy kernel.
The most general solutions of such  integral equations involve an
arbitrary number of polynomial parameters~\cite{Muskhelishvili,Omnes:1958hv}. 
Physically, the polynomial growth is limited by asymptotic conditions
on the amplitudes. In practice, however, the  system of equations is
valid only in the elastic scattering region, while the integrals run
%modif. figure
up to infinity. The polynomial part must thus be considered as a
parametrisation of the corrections to the effects of the integration
regions $s',t' \ge 1$ \Gevd over the solutions in the elastic scattering
region. For our purposes, we will consider here a four-parameter family of
solutions. The polynomial dependence is exhibited by introducing 
the Omn\`es functions, 
\be\lbl{omnesdef}
\Omega_I(w)=\exp \left[{w\over\pi}\int_{4\mpid}^\infty ds'\,
      {\delta_I(s')\over s'(s'-w)}\right]
\en
where $\delta_I$ is equal to the $S$ or $P$-wave $\pi\pi$ phase shift with
isospin $I$ in the elastic region. The functions $M_I$ are then expressed as
\bea\lbl{dispomnes}
&& M_0(w)=  \Omega_0(w)\left(\alpha_0 +\beta_0 w +\gamma_0 w^2 
 +  w^2 \hat{I}_0(w)\right)
\nonumber\\
&& M_1(w)=  \Omega_1(w)\left( \beta_1w +w\hat{I}_1(w)\right)
\nonumber\\
&& M_2(w)=  \Omega_2(w)\left(w^2 \hat{I}_2(w)\right)
\ena 
where
\be\lbl{Ihat}
\hat{I}_a(w)={1\over\pi}\int_{4\mpid}^\infty ds' {
  \sin\delta_a(s')\hat{M}_a(s')\over (s')^{2-n_a} 
(s'-w)\vert\Omega_a(s')\vert},\quad n_a=\delta_{1a} , 
\en
which ensure that the discontinuity eqs.~\rf{discMI} are obeyed.
The two subtraction constants which appear in the dispersive
representation~\rf{disprelMI} are simply related to the polynomial
parameters: $\tilde\alpha_0=\alpha_0$,
$\tilde\beta_0=\beta_0+\alpha_0\dot\Omega(0)  $. 
Plugging this representation into eq.~\rf{defMIhat} one obtains a set
of linear integral equations for the functions $\hat{M}_I$. As these are not
singular equations anymore, one may thus expect that, for given values
of the  parameters $\alpha_0$, $\beta_0$, $\gamma_0$, $\beta_1$, if a solution
exists for $\hat{M}_I$, it should be unique\cite{Anisovich:1996tx}.

The most natural idea for determining the  polynomial parameters
is by matching the KT amplitude with the amplitude computed in the
chiral expansion~\cite{Neveu:1970tn,Kambor:1995yc,Anisovich:1996tx} in
the region where the variables $s$, $t$, $u$ are small. 
More precisely, if $\widebar{M}(s,t,u)$ is the amplitude computed to
chiral order $p^N$, then the parameters $\alpha_0$, $\beta_0$,
$\beta_1$, $\gamma_0$ should be such that
\be
M(s,t,u)-\widebar{M}(s,t,u)=O(p^{N+2})\ .
\en
Considering the $N=4$ case, a first observation is that the
differences of the discontinuities in each of the component functions
$M_I$ are of chiral order $p^6$,
\be
\disc[M_I(w)-\widebar{M}_I(w)]=O(p^6)
\en
independently of the values of the polynomial parameters. 
This implies that the $O(p^4)$ parts of the differences 
$M_I(w)-\widebar{M}_I(w)$
must be polynomial. Imposing that the $O(p^4)$ parts of polynomial
expansions of the differences $M_I-\widebar{M}_I$ vanish  yields the
following four matching equations~\cite{Anisovich:1996tx}
\be\lbl{matchrels}
\ba{ll}
\alpha_0=& 9\,\big(\dfrac{1}{2}\widebar{M}''_2-\hat{I}_2\big)\,s_0^2 
        +3(\widebar{M}'_2-\widebar{M}_1)\,s_0
        +\widebar{M}_0+\dfrac{4}{3}\widebar{M}_2\\[0.25cm]
\beta_0=&\!\!\!-9\,\big(\dfrac{1}{2}\widebar{M}''_2-\hat{I}_2\big)\,s_0 
       + \widebar{M}'_0+3\widebar{M}_1-\dfrac{5}{3}\widebar{M}'_2
       -\Omega'_0\alpha_0\\[0.25cm]
\beta_1=&
\widebar{M}'_1+\dfrac{1}{2}\widebar{M}''_2-\hat{I}_1-\hat{I}_2\\[0.25cm] 
\gamma_0=& \dfrac{1}{2}\widebar{M}''_0+\dfrac{2}{3}\widebar{M}''_2
        -\hat{I}_0-\dfrac{4}{3}\hat{I}_2
        -\dfrac{1}{2}\Omega''_0\alpha_0-\Omega'_0\beta_0\ .
\ea\en
where the functions $\widebar{M}_a$, $\Omega_a$, $\hat{I}_a$  and
their derivatives are all to be taken at $w=0$. In order to solve the
set of eqs.~\rf{matchrels}, one must  keep in mind that the
integrals $\hat{I}_a$ carry an implicit linear dependence on the four
polynomial parameters. This dependence must be determined by using
four independent KT solutions in which one of the polynomial
parameters is set to one and the others to zero.

The  chiral NLO
amplitude $\widebar{M}(s,t,u)$ was first computed in
ref.~\cite{Gasser:1984pr}, it is given in 
ref.~\cite{Anisovich:1996tx} in a form which involves a single
$O(p^4)$ coupling constant, $L_3$. We will use here the
value $L_3=(-3.04\pm0.43) 10^{-3}$~\cite{Bijnens:2011tb}. 

\subsection{Numerical solutions and comparisons with the  $\eta\to 3\pi$
  data}\lblsec{5.4}
We have constructed numerical solutions of the set of KT equations by
iteration. The main differences with earlier
work\cite{Kambor:1995yc,Anisovich:1996tx} is that:  
\begin{itemize}
\item[a)] we
have used the kernel representations~\rf{kernelsproj} for performing
the angular integrations, which should be somewhat faster than the
integration over a complex contour method used previously,
\item[b)] The matching with ChPT was done via the  four matching
equations~\rf{matchrels}, which were solved with no approximations.  
\end{itemize}
%modif
In ref.~\cite{Anisovich:1996tx} only the last two matching equations
were implemented while the first two were replaced by imposing that
the amplitude $M(s,t,u)$ along the line $u=s$ has an Adler zero at the
same position and with the same slope as the chiral NLO amplitude.
We will see below (fig.~\fig{MAdler} ) that the first two matching
relations ensure essentially equivalent constraints on the Adler zero.

\begin{figure}[ht]
\bc
\includegraphics[width=0.7\linewidth]{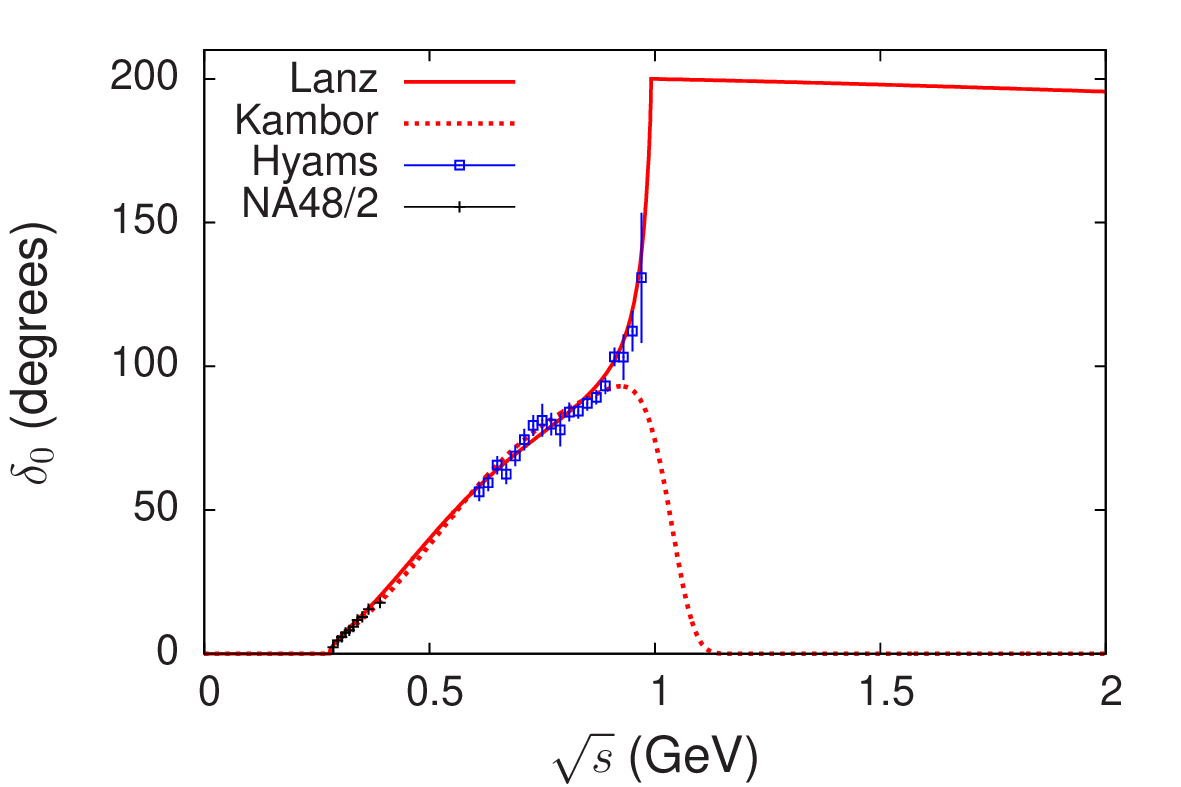}
\caption{\small Illustration of  $I=0$ phases used for computing the
  Omn\`es function $\Omega_0$ (eq.~\rf{omnesdef}) based on different
  assumptions in the inelastic region. The solid line is analogous to
  the phase used in ref.~\cite{Lanz:2011} and the dashed line is the
  phase used in ref.~\cite{Kambor:1995yc}. Also shown are some experimental
  $\pi\pi$ data from refs.~\cite{Hyams:1973zf,Batley:2010zza}. }
\label{fig:swave0}
\ec
\end{figure}

\begin{figure}[ht]
\bc
\includegraphics[width=0.7\linewidth]{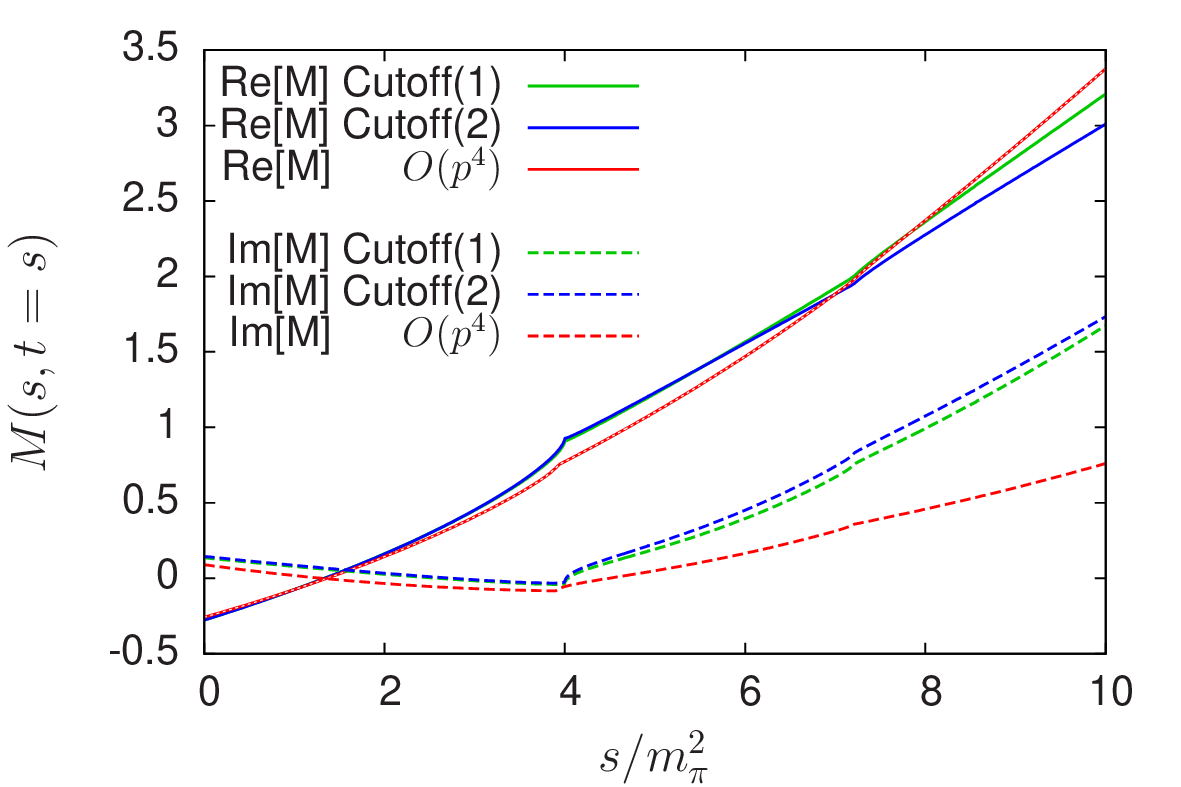}
\caption{\small Results for the amplitude $M(s,t)$ along the line
  $t=s$ obtained from solving the KT equations together with the four
  chiral matching relations~\rf{matchrels} using two different cutoff
  conditions for the phase $\delta_0$. Also shown is the chiral
  $O(p^4)$ amplitude.} 
\label{fig:MAdler}
\ec
\end{figure}

Concerning the phases $\delta_I(s)$ for $I=1$ and $I=2$, 
for which inelasticity sets in rather smoothly,
we take the phases to be equal to the corresponding $\pi\pi$
scattering phase shifts up to $\sqrt{s}_{cut}=1.4$ GeV and, for $s > s_{cut}$
interpolate to the asymptotic values $\delta_1(\infty)=\pi$ and
$\delta_2(\infty)=0$. 
\begin{table}[hbt]
\bc
\bt{ c|c|c|c|c}\hline\hline
\ & $\alpha_0$ & $\beta_0$ & $\beta_1$ & $\gamma_0$ \\ \hline
cutoff(1) &  $-0.60+i0.07$ & $15.7-i0.69$ &  $6.95+i0.40$ &  $-0.77+i0.86$   \\ 
cutoff(2) &  $-0.61+i0.08$ & $16.5-i0.88$ &  $6.89+i0.47$ &  $-26.5+i1.76$\\ 
fit       &  $-0.77-i0.02$ & $19.8-i0.17$ &  $4.75$ &  $-34.9$ \\
\hline\hline
\et
\caption{\small Influence of the cutoff conditions (see text)  for
the phase $\delta_0$ on the values of the polynomial parameters derived from
the matching equations (in appropriate powers of GeV). Also shown are
the results of fitting two parameters to the experimental
Dalitz plot results.
}
\lbltab{polyvalues}
\ec
\end{table}
In the case of the $I=0$ $S$-wave, inelasticity sets in sharply around
the $K\bar{K}$ threshold. We have employed two  different phase
choices in the inelastic region: a) the one used in
ref.~\cite{Kambor:1995yc} (which we call cutoff (1)): for $s$ larger
than $s_{cut}=(0.865)^2$ GeV$^2$, the phase is interpolated rapidly to
$\delta_0(\infty)=0$ and b)  a condition similar to that used in 
ref.~\cite{Lanz:2011}) (which we call cutoff (2)): for $s$ larger than
$4m_K^2$, the phase is interpolated slowly to
$\delta_0(\infty)=\pi$. These phases are illustrated in fig.~\fig{swave0}. 
When used in the matching relations~\rf{matchrels} these
different conditions lead to rather different values for some of the
polynomial parameters,\footnote{In particular, the simple estimate
given in ref.~\cite{Anisovich:1996tx}
$\gamma_0\simeq0$ is valid only if the cutoff is sufficiently small.},
reflecting differences in the values of the
integral $\hat{I}_0(0)$ as well as in the values of the Omn\`es function and
its derivatives at $s=0$ (see table~\Table{polyvalues}). However, the
matching conditions ensure that the complete KT amplitude depends only
moderately on the cutoff conditions. Fig.~\fig{MAdler} illustrates
some results, showing amplitudes along the line $t=s$ where an Adler
zero is present in the chiral amplitude.
%%%%%%%%%%%%%%%%%%
%%%%%%%%%%%%%%%%%% 

In order to assess the reliability of the $\eta\pi\to\pi\pi$ amplitude
resulting from KT solutions with ChPT matching, let us compare with
experimental results. From the  integrated decay rate of the charged
mode: $\Gamma_{\eta\to \pi^0\pi^-\pi^+}^{exp}=300\pm11$
eV~\cite{Beringer:1900zz}, one obtains for the central value of the
double quark mass ratio: $Q\simeq 21.6$ with the cutoff (1) condition
and $Q\simeq 21.5$ with the cutoff (2). This value is compatible with
the result of ref.~\cite{Kambor:1995yc} ($Q=21.6\pm1.3$, with NLO
matching) and slightly smaller than the one quoted in
ref.~\cite{Lanz:2011} ($Q=22.7$), based on the same formalism, but a
somewhat different implementation of the matching with ChPT. The
corresponding value of the quark mass ratio $\epsilon$, is
$\epsilon=1.32\cdot10^{-2}$. Using this value of $\epsilon$ in the
chiral expansion of $\fplus(0)$, eq.~\rf{fplus0NLO2},  on obtains a
result compatible with the one given in eq.~\rf{f+(0)exp}, derived
from experimental  data of $K^+_{l3}$, $K^0_{l3}$ decays. 
%

%%%%%%%%%%%%%%%%%%
%%%%%%%%%%%%%%%%%%
Precise measurements of the differential decay
distributions across the Dalitz plot have been performed recently for
both the charged~\cite{Ambrosino:2008ht} and the neutral decay
modes~\cite{Adolph:2008vn,Unverzagt:2008ny,Prakhov:2008ff,Ambrosinod:2010mj}. 
It is customary to represent these  differential distributions in terms of
a polynomial of two independent energy variables $X$, $Y$ (defined
such that $X^2+Y^2\le 1$ and  $X=Y=0$ corresponds to the center of the
Dalitz plot, where the three pions have equal kinetic energies, see
appendix~\sect{dalitzparam}) 
\be
{d^2\Gamma_c\over dX dY}(X,Y)={d^2\Gamma_c\over dX dY}(0,0)
\left(1+aY+bY^2+dX^2+fY^3+\cdots\right)
\en
for the charged mode, and
\be
{d^2\Gamma_n\over dX dY}(X,Y)={d^2\Gamma_n\over dX dY}(0,0)
\left(1+\alpha(X^2+Y^2)+\cdots\right)
\en
for the neutral mode. 
The experimental values of the Dalitz plot parameters $a$, $b$, $d$,
$f$ and $\alpha$ are shown in table~\Table{dalitz} together with
results corresponding to KT solutions amplitudes. Implementation of
rescattering effects via the KT equations leads to significantly
improved results with respect to the simple use of the chiral NLO
amplitude (see~\cite{Gasser:1984pr})
but the two Dalitz parameters $b$ and $d$ are still predicted to
be  too large. 
\begin{table}[tb]
\bc
\bt{ c|c|c|c|c }\hline\hline
\TT Dalitz par. & experimental  & cutoff(1) & cutoff(2) & Fit \\ \hline
\TT a &$-1.090\pm0.005^{+0.008}_{-0.019}$ & $-1.171$ &  $-1.125$  & $-1.062$\\
\TT b &$ 0.124\pm0.006\pm 0.010    $ & $0.260$  &  $0.196$  & $0.163$ \\
\TT d &$ 0.057\pm0.006^{+0.007}_{-0.016}$ & $0.083 $ &   $0.082$ & $0.067$\\
\TT f &$ 0.14 \pm0.01 \pm 0.02$      & $0.074 $ &   $0.100$  & $0.102$\\ \hline
$\alpha$ & $-0.0315\pm 0.0015$       & $-0.0127$&  $-0.0260$ & $-0.0336$ \\   \hline\hline
\et
\caption{\small  Comparison between the experimental values of the
  Dalitz plot parameters for $\eta\to \pi^0\pi^+\pi^-$ and
  $\eta\to3\pi^0$ and the predictions of the KT solution amplitudes
  with NLO matching and two different cutoff conditions (see
  table~\Table{polyvalues}). The last column shows the result of a KT
  solution where part of the polynomial parameters are fitted to the data.} 
\lbltab{dalitz}
\ec
\end{table}

%modif
This discrepancy between the theoretical amplitude and experiment
indicates that further effects need to be taken into account. These
could be either chiral $O(p^6)$ effects at the level of the matching
equations or further rescattering effects. In this respect, one sees
clearly from table~\Table{dalitz} that the $\pi\pi$ phase choice 
which does include the $f_0(980)$ resonance leads to better results for
the Dalitz plot parameters than the choice which does not. It is then
not unlikely that the $a_0(980)$ resonance should be taken into
account as well. It is also worth noting that preliminary results from
analysis of new data sets by KLOE and WASA have been presented
(see~\cite{Amaryan:2013eja}, p.16) which go in the direction of
improving the agreement with the theoretical predictions.

For our present purposes, in addition to the KT amplitudes which obey
the ChPT matching relations, we construct an amplitude which
reproduces more closely the experimental results on the Dalitz
plot. In order to do so, we allow the two polynomial parameters
$\beta_1$, $\gamma_0$ to vary freely (still assuming their imaginary
parts to be negligible) and use them as fit parameters. The last two
polynomial parameters $\alpha_0$, $\beta_0$ are then fixed from the
two conditions: 1) that the amplitude reproduces the position of the
Adler zero $s_A$ of the NLO amplitude and 2) that the central value of
the quark mass double ratio $Q$ from lattice QCD (the recent FLAG
review~\cite{Aoki:2013ldr} gives $Q^{latt}=22.6(7)(6)$ from $N_f=2+1$
simulations) is reproduced. The results from this amplitude for Dalitz
plot parameters are displayed in the last column of
table~\Table{dalitz} and the corresponding polynomial parameters are
shown on the last line of table~\Table{polyvalues}.

%modif in figure: pw finite when s -> 4mpi2(+)
%%%%%%%%%%%%%%%%%%
\begin{figure}[htp]
\bc
\includegraphics[width=0.7\linewidth]{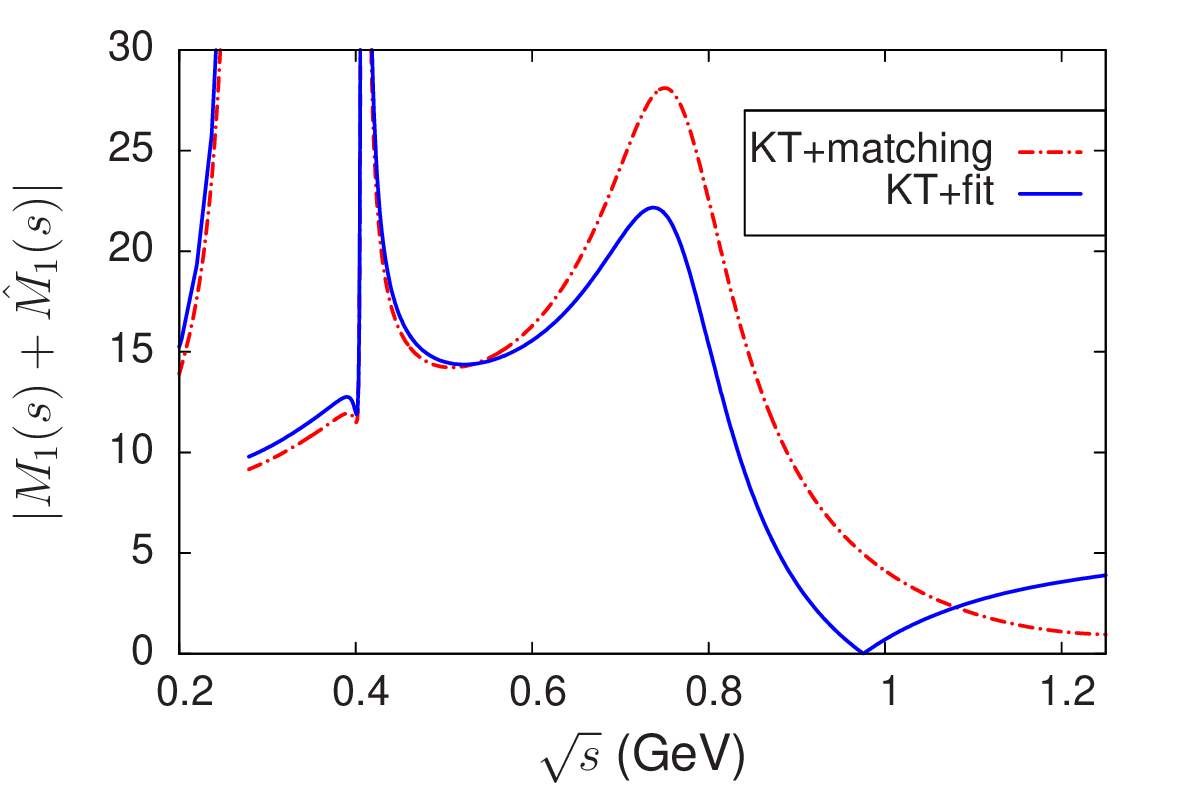}
\caption{\small Modulus of function $M_1+\hat{M}_1$ obtained from KT solutions
  with two sets of polynomial parameters. Dash-dotted curve:
  parameters from $O(p^4)$ matching (cutoff (2)), solid curve: parameters from
fit. }
\label{fig:M1}
\ec
\end{figure}
%%%%%%%%%%%%%%%%%% 
\subsection{Results for $\fplus$}
In the elastic scattering region, the discontinuity of $\fplus(s)$ was
given in eq.~\rf{unitfplus} in terms of the pion form factor $F_V^\pi$
and the $l=1$ projection of the $\eta\pi^+\to\pi^0\pi^+$ amplitude.
From the decomposition theorem, this projected amplitude may be
expressed in terms of $M_1+\hat{M}_1$ (eq.~\rf{discfplusM1}). We can
now calculate this quantity from our solutions of the KT
equations. Fig.~\fig{M1} shows the numerical results for the modulus
of $M_1+\hat{M}_1$. It illustrates the strong sensitivity to the
choice of the polynomial parameters. The solution corresponding to
fitted parameters (last line in table~\Table{polyvalues}) has a
significantly smaller resonance peak, which is related to the smaller
size of the parameter $\beta_1$.
%modif here: more details on Fv

Concerning the pion form factor, we used the
experimental measurements of $\vert F_V^\pi \vert$ from $\tau$ decays
(which provide exactly the same form factor as needed here) by the Belle
collaboration~\cite{Fujikawa:2008ma}. The measurement covers the
energy range $0.297 \le \sqrt{s} \le 1.255$ GeV which is essentially
adequate for our purposes. We relied on the fit performed in
ref.~\cite{Fujikawa:2008ma} in terms of Gounaris-Sakurai (GS)
functions~\cite{Gounaris:1968mw} for performing an extrapolation of
$\vert F_V^\pi \vert$ in the small energy region down to the threshold
and performing the numerical integrations. For the phase
$\delta_V^\pi$, we assumed elastic unitarity to hold below 1 GeV and
thus took $\delta_V^\pi$ to be equal to the $l=1$ $\pi\pi$ phase shift
in accordance with Watson's theorem. Above 1 GeV, we use the phase as
predicted by the GS parametrisation\footnote{Below 1 GeV, the phase
$\delta_V^\pi$ produced by the GS parametrisation and the $\pi\pi$
phase shift are quite close, differing by $5-10\%$, except at low
energy, $\sqrt{s} < 0.5$ GeV, where the difference is more
significant.}.

Asymptotically, from the usual QCD-based
arguments~\cite{Duncan:1979hi,Lepage:1980fj}, one expects the form
factor $\fplus(s)$ to behave as $1/s\log(s)$, it should thus obey a
convergent dispersion relation.  
We will actually use DR's for $\fplus(s)/s^N$ in order to suppress the
contribution from the integration region above 1 GeV. In practice,
we will use $N=1$ or $N=2$ and check the stability of the result. For
example, with $N=2$, the DR reads
\be\lbl{dispfplus}
\fplus(s)=\fplus(0) + s\dotfplus(0)+{s^2\over\pi}\int_{4\mpid}^\infty
ds'\, { \disc[ \fplus(s')]\over (s')^2  (s'-s)}\ .
\en
The values of $\fplus$ and its derivative at $s=0$ from NLO ChPT were
given in sec.~\sect{resultschpt}. The value of the quark mass ratio
$\epsilon$ used in eq.~\rf{dotf+(0)exp} was deduced from the quark
mass double ratio $Q$ corresponding to the KT amplitude used
(i.e. either $Q=21.5$ with matched polynomial parameters or $Q=22.6$
with fitted parameters) and using the central value of the result
given in the FLAG review~\cite{Aoki:2013ldr}: $2m_s/(m_u+m_d)=27.46(15)(41)$.

%%%%%%%%%%%%%%%%%%
\begin{figure}[hb]
\bc
\includegraphics[width=0.7\linewidth]{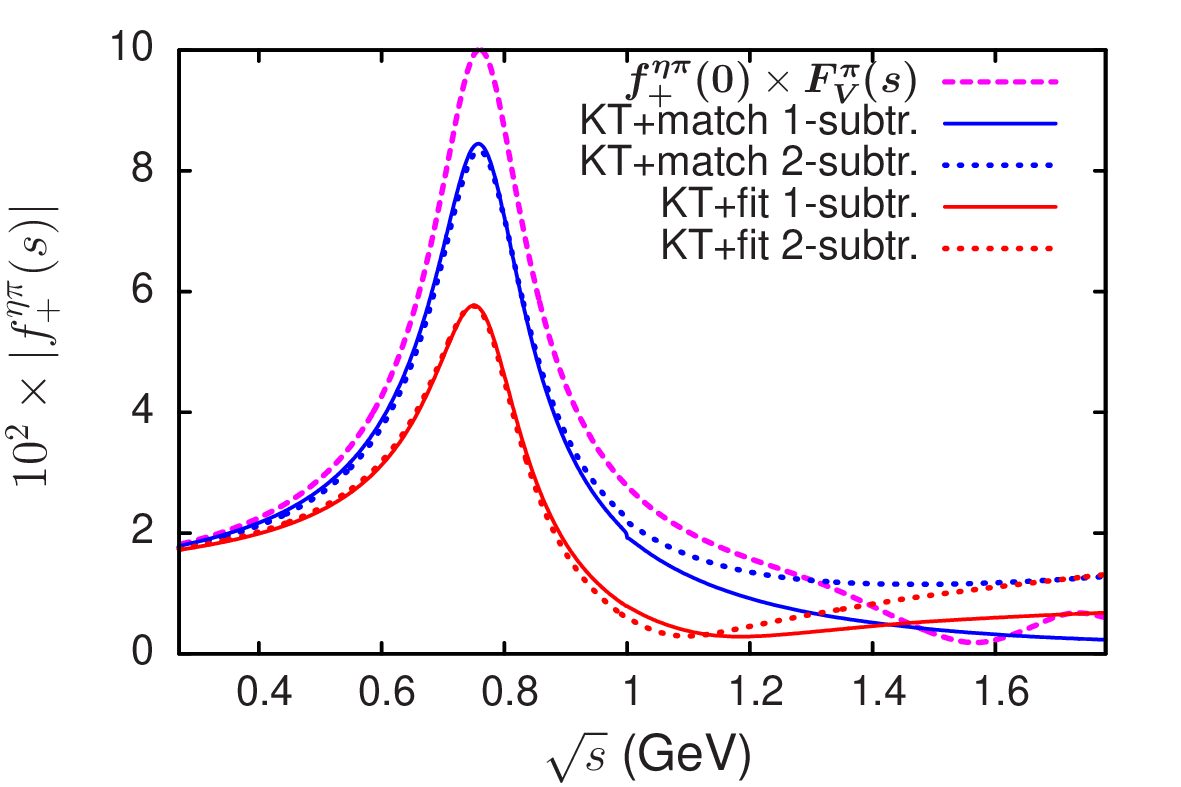}
\caption{\small Dispersive calculations of the vector form factor
  $\fplus$. The upper and lower solid lines correspond to
  once-subtracted dispersion dispersion relations using KT amplitudes
  with matched  and with fitted parameters respectively.   The dotted
  lines correspond to twice subtracted dispersion relations. The pion
  vector form factor normalised to $\fplus(0)$ at $s=0$ is also shown
  for comparison (dashed line).} 
\label{fig:fplus}
\ec
\end{figure}
%%%%%%%%%%%%%%%%%% 
The results obtained using these dispersion relations  together with
$\eta\pi\to\pi\pi$ amplitudes from KT equations and also using the
parametrisation given by the Belle
collaboration~\cite{Fujikawa:2008ma} for the pion form factor are
shown in fig.~\fig{fplus}. The figure shows that the DR's with $N=1$
or $N=2$ yield very similar results in the region $\sqrt{s}\le 1$
GeV. As one can expect from the behaviour of $M_1+\hat{M}_1$ (see
fig.~\fig{M1}) the KT amplitude with matched polynomial parameters
gives rise to a larger $\rho$ peak than the one with the fitted
parameters and the peak is more similar in  shape to that of the pion
vector form factor. 

\section{Dispersive estimate of $\fzero$}
In this section we provide a qualitative estimate of the scalar form
factor, assuming that there should be  analogies between the $\pi\eta$
scattering amplitude and the well-known $\pi\pi$, $\pi K$ scattering
amplitudes and also between $\fzero$ and the $\pi K$ scalar form factor.

\subsection{Phase dispersive representation}
The unitarity relations obeyed by the scalar form factor $\fzero$, associated
with the the two-body channels $\pi\pi$, $\pi\eta$ and $K\Kbar$ were
written in sec.~\sect{unitfzero}. Below the $K\Kbar$ threshold, the
contribution from $\pi\eta$ largely dominates since $\pi\pi$ is
comparatively suppressed by isospin symmetry. It is then convenient to
write the form factor as a phase dispersive representation\footnote{We
  make the usual assumption that no nearby complex zeros are present.},
\be\lbl{fzerodisp}
\fzero(s)=\fzero(0)\times\exp\bigg[ \zeta s
  +{s^2\over\pi}\int_{(\meta+\mpi)^2}^\infty {\phi^{\eta\pi}(s')\over
    (s')^2(s'-s)} ds'\bigg]
\en
where $\phi^{\eta\pi}$ is the phase of the form factor and
$\zeta=\dotfzero(0)/\fzero(0)$. 
The representation~\rf{fzerodisp}
uses two subtractions in order to suppress, as much as possible, the
contributions from the higher-energy regions. The values of the form
factor and its derivative at $s=0$ can be taken from ChPT at $O(p^4)$,
see sec.~\sect{resultschpt}. Below the $K\Kbar$ threshold, $\eta\pi$
scattering can be assumed to be essentially elastic such that, in this
region, the form factor phase can be identified with the $l=0$ elastic
scattering phase shift $\delta_0^{\eta\pi}$ by Watson's theorem. 

The $S$-wave $\eta\pi$ scattering phase shift has not yet been directly
measured, but detailed experimental information exists on the
properties of the scalar resonances $a_0(980)$ and
$a_0(1450)$. Furthermore, chiral symmetry constrains the phase shift
to be very small at low energy~\cite{Bernard:1991xb}. For
definiteness, we will make use of the $\eta\pi$-scattering model
proposed in ref.~\cite{Black:1999dx}, which  encodes
these various pieces of information in a simple way. It uses
constraints on $\eta\pi$ scattering derived from the
$\eta'\to\eta\pi\pi$ decay amplitude~\cite{Fariborz:1999gr}. The
$\eta\pi$ amplitude is written in the following form
\be
{\cal T}^{\eta\pi}(s,t,u)= T_{CA} + A_0(s)+ A_0(u) + F_0(t)
\en
where the first term is the constant current algebra contribution,
\be
T_{CA}= {\mpid\over 3\fpid}(\cos\delta-\sqrt2\sin\delta)^2
\en
(accounting for $\eta-\eta'$ mixing, $\delta$ being the corresponding
octet-singlet angle\footnote{A different convention for $F_\pi$ and for
  the mixing angle was used in ref.~\cite{Black:1999dx}.}). The
function $A_0$ represents a sum over tree-level amplitudes 
associated with the $a_0(980)$, $a_0(1450)$ resonances and $F_0$ is a
similar sum involving  the two isoscalar resonances $\sigma(600)$ and
$f_0(980)$. These amplitudes are computed from a resonance chiral
Lagrangian and thus behave as $O(p^4)$ at small values of the
Mandelstam variables\footnote{This model predicts $a_0\simeq 
%modif: note Kubis normaliz. different from BKM
4.1\times10^{-2}$ for the $l=0$ scattering length. This  value
is somewhat larger than the NLO ChPT low-energy theorem
result~\cite{Kubis:2009sb} $a_0=(-0.02\pm 0.77)\times10^{-2}$. It
can possibly be accommodated in schemes where higher order effects
associated with OZI violations are included~\cite{Kolesar:2008jr}.}.
We used the set of resonance parameters from
eqs. (4.7), (A17), (A1) of ref.~\cite{Black:1999dx}. The partial-wave
amplitudes being given by 
\be
t^{\eta\pi}_l(s)={\lambetapi{s}\over32\pi s}\int_{-1}^1 dz\,
P_l(z){\cal T}^{\eta\pi}(s,t,u)
\en
we define the phase shift from the ansatz
\be
\sin(2\delta_l^{\eta\pi})= {2\re{ t^{\eta\pi}_l(s) }\over
\vert 1+2i t^{\eta\pi}_l(s) \vert\ }\ , 
\en
which applies also in the inelastic scattering region. 
The $l=0$ phase shift from this model is shown in fig.~\fig{etapiphase}.
It is in qualitatively good 
agreement with other  approaches which have been proposed like the
chiral unitary approach~\cite{Oller:1998hw}. It is also in agreement
with one of the models used in ref.~\cite{Achasov:2010kk} and probed
against the recent high-statistics measurements of
$\gamma\gamma\to\eta\pi$ scattering by the Belle
collaboration~\cite{Uehara:2009cf}. 
%%%%%%%%%%%%%%%%%%
\begin{figure}[ht]
\bc
\includegraphics[width=0.7\linewidth]{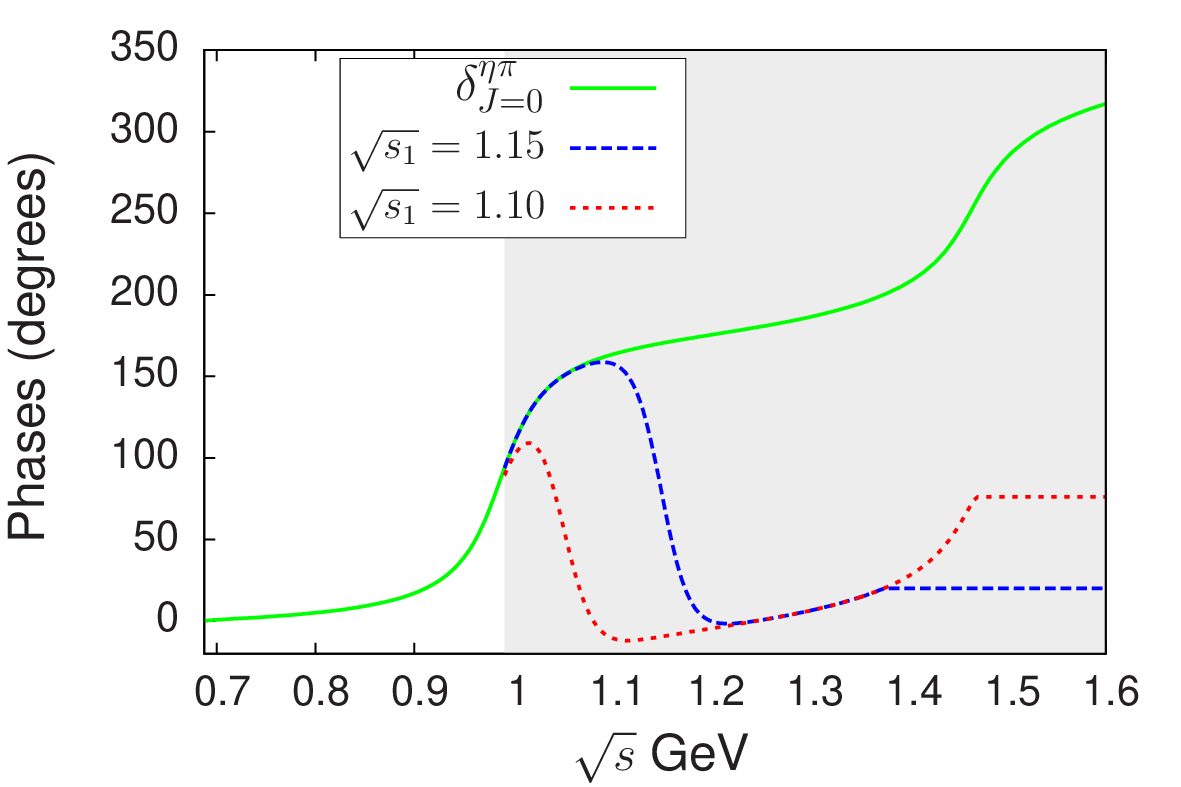}
\caption{\small  The solid line is the $l=0$ $\pi\eta$ scattering
  phase shift obtained from the model of ref.~\cite{Black:1999dx}, the
  shaded area indicates the region where scattering is inelastic. The
  dashed lines shows  the scalar form factor phase $\phi^{\eta\pi}$
  for two values of $s_1$ (see text).} 
\lblfig{etapiphase}
\ec
\end{figure}
%%%%%%%%%%%%%%%%%%

Above the $K\Kbar$ threshold, $\eta\pi$ scattering becomes inelastic
under the effect, mainly, of the two-body channels $K\Kbar$ and
$\eta'\pi$. The phase $\phi^{\eta\pi}$ must then differ from
$\delta_0^{\eta\pi}$. A global constraint on $\phi^{\eta\pi}$ arises from
imposing that the form factor, as given from eq.~\rf{fzerodisp},
exhibits no exponential divergence asymptotically. This gives rise to
a sum rule
\be\lbl{phietapisumrule}
\zeta\equiv {\dotfzero(0)\over\fzero(0)}= {1\over\pi} \int_{(\meta+\mpi)^2}^\infty
{\phi^{\eta\pi}(s')\over (s')^2}\,ds'\ .
\en
Using the chiral expansion results for $\zeta$ (see sec.~\sect{scalarffp4}),
the sum rule indicates that one should have $\phi^{\eta\pi} <<
\delta_0^{\eta\pi}$ in the inelastic region. In order to estimate more
precisely how $\phi^{\eta\pi}$ behaves, one may rely on
an analogy with the phase of the $K\pi$ scalar form
factor. In that case, there is enough experimental information on the
elastic as well as the leading two-body inelastic $T$-matrix elements,
such that the form factor can be deduced from solving a set of
Muskhelishvili-Omn\`es equations. The analysis performed in
ref.~\cite{Jamin:2001zq} shows that the form factor phase displays a sharp
drop shortly after the onset of the leading inelastic
channel\footnote{Only the modulus of the form factor is actually displayed in
  ref.~\cite{Jamin:2001zq}. One can find both the modulus and the corresponding
  phase shown in Fig.~1 of ref.~\cite{ElBennich:2009da}.}.
A similar behaviour has been also observed in the case of the
$\pi\pi$ scalar form factor associated with the $\bar{u}u+\bar{d}d$
operator~\cite{Ananthanarayan:2004xy}. The authors argue that the
presence of this phase drop is  necessary in order to reproduce the
correct value of the pion scalar radius via a sum rule analogous to
eq.~\rf{phietapisumrule}. We then propose the following simple model for the
phase $\phi^{\eta\pi}$ assuming a fast decrease at $s=s_1$ and a 
constant value for $s > s_2$ with $4\mkd < s_1 < s_2 $: 
\be\lbl{phietapimodel}
\ba{ll}
s \le s_2:         & \phi^{\eta\pi}(s)= \delta_0^{\eta\pi}(s)-\pi\theta(s-s_1) \\
s > s_2 :         &  \phi^{\eta\pi}(s)= \phi^{\eta\pi}(s_2)\\
\ea
\en
(a slight smoothing of the $\theta$ function is implemented in
practice). For each value of $s_1$, the value of $s_2$ is determined
such that the sum rule~\rf{phietapisumrule} is exactly
satisfied. Fig.~\fig{etapiphase} illustrates the behaviour of
$\phi^{\eta\pi}$ with $\sqrt{s_1}=1.05,\ 1.15$ GeV. In this
model, $s_1$ is bounded from above: $\sqrt{s_1} < 1.2$ GeV for otherwise, 
it is not possible to satisfy the sum rule~\rf{phietapisumrule}.  
\subsection{Results for $\fzero$}
%%%%%%%%%%%%%%%%%%
\begin{figure}[ht]
\bc
\includegraphics[width=0.7\linewidth]{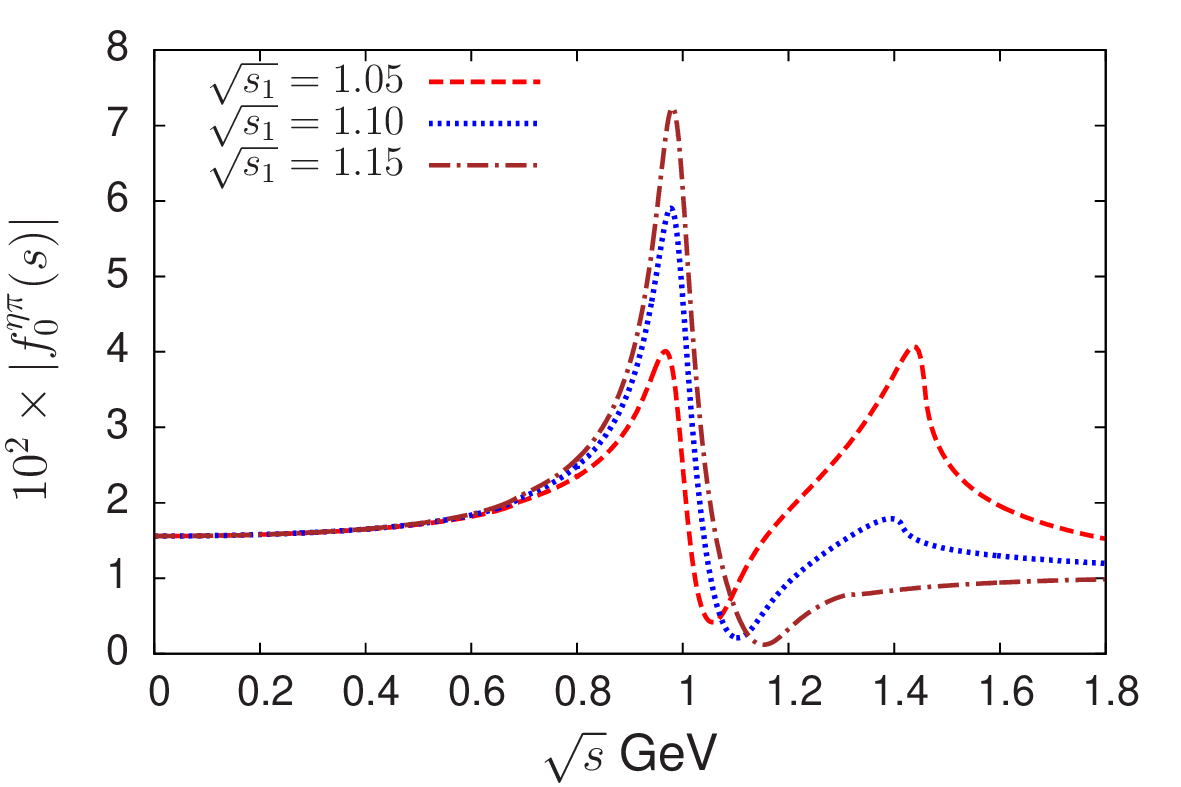}
\caption{\small Modulus of the scalar form factor from the phase dispersive
  representation~\rf{fzerodisp} and the phase model described in the
  text, for several values of the dip position $s_1$.}
\label{fig:fzero}
\ec
\end{figure}
%%%%%%%%%%%%%%%%%% 
Using the phase $\phi^{\eta\pi}$ as described above, one can compute
the scalar form factor from the dispersive representation~\rf{fzerodisp}. 
Results are shown in fig~\fig{fzero} for several values of $s_1$.  
The point $s_1$, where the phase drops, corresponds to a dip in the
modulus of the form factor. Obviously, if the dip is located very close to the
$K\bar{K}$ threshold, the peak of the $a_0(980)$ resonance is strongly
reduced. This corresponds, in this approach, to a reduced coupling of
the resonance to the $\bar{u}d$ scalar operator and thus to an
exotic nature of the  $a_0(980)$.
One may formally define a coupling constant from the matrix element of
the vector current involving the $a_0$ state~\cite{Narison:1988xi}
\be
\braque{0\vert \Vmud\vert a_0^-(p)} =\sqrt2 F_{a_0} p^\mu\ .
\en 
In a dispersive approach, it is in principle possible to identify
such a coupling constant from the residue of the $a_0$ resonance pole
of the scalar form factor  on the second Riemann sheet. In a simpler
way, one may obtain an estimate by matching the shape of the
dispersive form factor with a Breit-Wigner-type shape. For this
purpose, we have used the  chiral resonance approach of
ref.~\cite{Neufeld:1994eg} in which one can vary the value of
$F_{a_0}$ (via that of an $O(p^6)$ chiral coupling constant, $d_r$)
while the value of the form factor at the origin is kept fixed.  In
this manner, from the phase dispersive representation, with the
largest allowed value of the dip parameter, we find 
\be\lbl{Fa0}
F_{a_0}\simeq 0.62\ \hbox{MeV}\ 
\en
which thus represents an upper bound for this coupling in the present
model. For comparison, based on QCD sum rules, values in the range
$[0.8-1.6]$ MeV have been quoted~\cite{Maltman:1999jn,Narison:1988xi}.
In the same framework, it was found in ref.~\cite{Elias:1998bq} that,
on the contrary, the $a_0$ coupling should be strongly suppressed,
while estimates based on the bag model give a range of $[0.2-2.0]$
MeV~\cite{Fajfer:1988fe}. A result from a lattice QCD simulation with
$N_f=2$ dynamical quarks has been given~\cite{McNeile:2006nv}, which
corresponds to $F_{a_0}=[0.8-0.9]$ MeV.

%%%%%%%%%%%%%%%%%%
\begin{figure}[ht]
\bc
\includegraphics[width=0.7\linewidth]{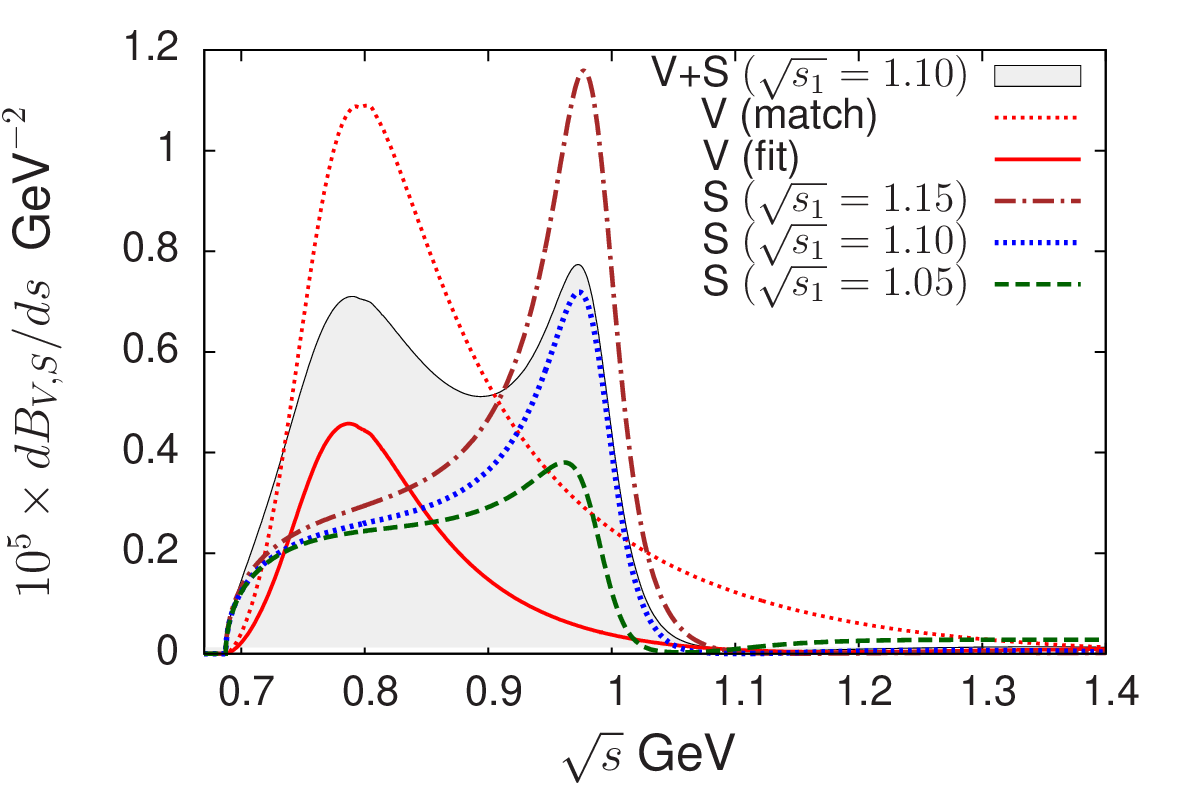}
\caption{\small Contributions to the invariant mass distribution of
  the $\tau\to\eta\pi\nu$ branching fraction from the vector (V) and the
  scalar (S) form factors. For the vector contribution, the upper
  dotted curve corresponds to polynomial parameters from ChPT matching
  and the lower solid curve to  parameters fitted to experiment. The scalar
  contribution is shown for several values of the dip parameter $s_1$.
The shaded area shows a plausible central value for the  sum of the
two contributions .}  
\label{fig:etapispectral}
\ec
\end{figure}
%%%%%%%%%%%%%%%%%%
\section{Application to the $\tau\to \eta\pi\nu$ and $\eta_{l3}$ decays}
We can now compute the contributions from the vector and scalar form
factors to the differential decay width of the $\tau$ into
$\eta\pi\nu$ (the relevant formula was given in  eq.~\rf{differtaudecay}). 
Fig.~\fig{etapispectral} shows the contribution from the vector form
factor calculated from a KT amplitude with fitted polynomial
parameters (lower curve) and that associated with parameters determined
from ChPT matching (upper curve). 
The vector contribution is somewhat suppressed here by the kinematics but
our calculations suggest that it should lead to a clearly visible
$\rho$-meson peak.   
The contribution to the differential decay width from the scalar form
factor is shown for three different values of the dip $s_1$.

The corresponding values for the integrated branching fraction of
the $\tau\to \eta\pi\nu$ mode are given in
table~\Table{branchings} and compared with some former results found in the
literature. We quote here a plausible central value only, which
corresponds to $\sqrt{s_1}=1.10$ GeV for the scalar form factor and to
fitted polynomial parameters for the vector form factor. It is
difficult to precisely evaluate the error, in particular in the case
of the scalar form factor, because of the various assumptions and
model dependence involved. 
A plausible guess however, in our approach,  is that the scalar
contribution to the branching fraction should lie in the
range\footnote{Varying only the position of the dip parameter $s_1$
yields a range $B_S=[0.17-0.30]\cdot10^{-5}$.}
$0.1\cdot10^{-5} \le B_S \le 0.6\cdot10^{-5}$. This tends to be
smaller than most  previous estimates which are often based on simple
scalar-dominance models.
Conversely, in the dispersive approach, it seems difficult to
accommodate a value for $B_S$ as small as quoted in
ref.~\cite{Volkov:2012be} even if the position of the dip $s_1$
is very close to the $K\Kbar$ threshold, because of the contribution
from below the resonance region. 

In the case of the vector contribution, in view of some possible shifts
in the experimental value of the $\eta\to3\pi$ Dalitz plot parameters
(see~\cite{Amaryan:2013eja}, p.16), a plausible range for the branching
fraction should be: $0.10\cdot10^{-5}\le B_V\le 0.40\times 10^{-5}$. 
%
%%%%%%%%%%%%%%%table of branching fractions
\begin{table}[hbt]
\bc
\bt{cccl}\hline\hline 
\TT $10^5\times B_V$ &  $10^5\times B_S$  & $10^5\times B_{V+S}$
&\qquad ref.   \\ \hline 
\TT 0.25 & 1.60 & 1.85 & Tisserant, Truong (1982)\cite{Tisserant:1982fc} \\
 0.12 & 1.38 & 1.50 &   Pich (1987)      \cite{Pich:1987qq} \\
 0.15 & 1.06 & 1.21 &   Neufeld, Rupertsberger(1995)\cite{Neufeld:1994eg}\\
 0.36 & 1.00 & 1.36 & Nussinov, Soffer (2008)\cite{Nussinov:2008gx}\\
 $[$0.2-0.6$]$ & $[$0.2-2.3$]$ & $[$0.4-2.9$]$ &  Paver, Riazuddin
 (2010)\cite{Paver:2010mz} \\ 
0.44           &  0.04     & 0.48    & Volkov,Kostunin
(2012)\cite{Volkov:2012be} \\ \hline 
0.13  & 0.20  &  0.33  & Present work \\  \hline\hline  
\et
\caption{\small  Theoretical estimates of the branching fraction of the
  $\tau\to\eta\pi\nu$ mode, showing the separate contribution from the vector
  and scalar form factors. The central values from the approach used
  here are shown on the last line.} 
\lbltab{branchings}
\ec
\end{table}

Finally, for the $\eta_{l3}$ decays, we find the following central values for
the branching fractions (adding the two charge modes)
\be
\ba{l}
B_{\eta\to\pi^+ e^-\nu+c.c.} \simeq 1.40\times 10^{-13} \\
B_{\eta\to\pi^+ \mu^-\nu+c.c.} =1.02\times 10^{-13} \\
\ea
\en
The results, in this case, are practically identical to those computed
with the chiral NLO form factors\cite{Neufeld:1994eg,Scora:1995sj}. An
experimental upper bound on the $\eta_{e3}$ mode branching fraction
has been obtained recently by the BESIII~\cite{Ablikim:2012vn}
collaboration
\be
B_{\eta\to\pi^+ e^-\nu+ c.c.} <  1.7\times 10^{-4}\ .
\en
%modif: BF given 

%%%%%%%%%%%%%%
\section{Conclusions}
In this work, we have reconsidered the $\eta\pi$ isospin-violating
vector and scalar form factors and the related energy distribution in
the second-class $\tau\to \eta\pi\nu$ decay which should be measurable
at future $B$ or $\tau$-charm factories.
We have started  from the NLO ChPT results for $\fplus(0)$ and for the
derivatives $\dotfplus(0)$, $\dotfzero(0)$. In particular, for
$\fplus(0)$ a relation was established~\cite{Neufeld:1994eg} with the
$K^+_{l3}$ and $K^0_{l3}$ decays which can now be used thanks to
recent experimental progress~\cite{Antonelli10}. These results at
$s=0$ could be checked, in principle, in lattice QCD simulations
including isospin violation. 

In order to evaluate the form factors in the resonance regions we further
relied on their analyticity properties. We argued that these should be
the same as in the more familiar cases of the $\pi\pi$ or the $\pi K$
form factors, i.e., no anomalous threshold should be present despite
the fact that the $\eta$ meson is unstable. Its 
instability only generates  a few technical complications in the case of
$\fplus$: the discontinuity along the unitarity cut is complex and,
furthermore, it displays a divergence at the pseudothreshold
$s=(\meta-\mpi)^2$. 

Below 1 GeV, the essential contribution to the discontinuity is
proportional to the $\eta\pi  \to\pi\pi$ amplitude, projected on the
$P$-wave. We constructed a four-parameter family of solutions of the
Khuri-Treiman equations which we  use in the dispersion relation
for $\fplus$. The shape of the vector form factor, in particular the
$\rho$-meson peak, is then correlated with the Dalitz plot parameters
of the $\eta\to 3\pi$ amplitude (in particular, the parameter
$d$). Upon using the recent experimental constraints for the Dalitz
plot, we find the $\rho$-meson peak to be suppressed as compared with
earlier evaluations and that its shape differs from the naive vector
dominance model.

In the case of the scalar form factor $\fzero$, we used a phase
dispersive representation. For the $\eta\pi$ scattering phase shift in
the elastic region, we relied on the $\eta\pi$ scattering model
proposed in ref.~\cite{Black:1999dx}. 
This model should be reasonable, at the
qualitative level, but it is clear that there is much to be improved
on our knowledge of $\eta\pi$ scattering. Again in this case,  
lattice QCD simulations which are making steady progress 
in evaluating meson-meson interactions (see~\cite{Lang:2012sv} for
recent work), could provide unique information, e.g., on the value of
the scattering length.  

At energies above the $K\Kbar$
threshold, we argued that a plausible behaviour for the phase is that
it should display a sharp fall-off (which corresponds to a dip in the
modulus), by analogy with the cases of the $\pi\pi$ or the $\pi K$
scalar form factors, where the corresponding phase can be generated from
dynamical models. In this approach the exotic (non-exotic) nature of
the $a_0(980)$ resonance corresponds to the dip being situated close
(far) from the resonance position. 
This feature of the phase can then be used in
association with  a global constraint from a sum rule, which relates
the integral over the phase to the logarithmic derivative of the form
factor at the origin.  
Varying the position of the dip generates the main source of
uncertainty in this approach. The sum rule  restricts the range of
variation of the dip but the uncertainties still remain much larger than the
$20\%$ level required to make the $\tau\to\eta\pi\nu$ process
competitive for constraining the parameters of particle physics models
involving charged Higgs bosons.

\section*{Acknowledgements}
We would like to thank Helmut Neufeld for correspondence and Emi Kou
for discussions. 
This work is supported in part by the European Community-Research
Infrastructure Integrating Activity "Study of Strongly Integrating
Matter" (acronym HadronPhysics3, Grant Agreement Nr 283286) under the
Seventh Framework Programme of EU.
%\newpage
%\section*{Appendices}
\appendix
\section{Angular projection kernels}\lblsec{kernels}
In sec.~\sect{singmhat} the angular integrals of the functions
$M_I(t(s,z))$ were expressed in terms of kernels. The kernels needed
for $I=1$, firstly, have the following expression
\bea\lbl{PKernels}
&& P^{(0)}(t',s)={1\over t'} + L(t',s)\nonumber\\
&& P^{(1)}(t',s)={2\over\kappa(s)}+{(2t'+s-3s_0)\over \kappa(s)}\, L(t',s)
\nonumber\\
&& P^{(2)}(t',s)={1\over3 t'}+{2(2t'+s-3s_0)\over\kappa(s)^2}
+{(2t'+s-3s_0)^2\over\kappa(s)^2}\,L(t',s)\ ,
\ena 
where the logarithmic function $L(t',s)$ was given in
eq.~\rf{basiclog}. For $I=0,2$ the two kernels which are needed read
\bea\lbl{KKernels}
&& K^{(0)}(t',s)= {3s_0-s\over 2 (t')^2} + P^{(0)}(t',s)\nonumber\\
&& K^{(1)}(t',s)= {\kappa(s)\over 6(t')^2} +  P^{(1)}(t',s)
\ena

\section{Verification of the $O(p^4)$ discontinuities}\lblsec{verifdiscp4}
Let us verify here, using the general unitarity formulae for
$\fplus(s)$ and $\fzero(s)$ given in secs.~\sect{unitfplus} and
\sect{unitfzero} that one reproduces the $O(p^4)$ results. For this
purpose, one must use the chiral expansions of the form factors and those of
the scattering amplitudes which appear in the unitarity relations at
order $p^2$. The expressions for the three relevant scattering
amplitudes at $O(p^2)$ are
\bea
&& T_{\eta\piplus\to\pizero\piplus}(s,t,u)=-\epsilon\,{3t-4\mpid\over3F_0^2},
\nonumber\\
&& T_{\eta\piplus\to\eta\piplus}(s,t,u)={\mpid\over 3F_0^2}
\nonumber\\
&&T_{\eta\piplus\to\kzerob\kplus}(s,t,u)={\sqrt6(3s -4\mkd)\over12 F_0^2}
+\sqrt2\epsilon{t-u\over 4F_0^2}\ .            
\ena
The unitarity relation for $\fplus$ involves the $l=1$ partial-wave
projections of these amplitudes which read
\bea
&& {1\over2}\int_{-1}^{1} dz z\, T_{\eta\piplus\to \pizero\piplus}=
-\epsilon\,\sqrt{{s-4\mpid\over s}}\,
{\sqrt{\lambda_{\eta\pi}(s)}\over 6 F_0^2},\ 
\nonumber\\
&& {1\over2}\int_{-1}^{1} dz z\, T_{\eta\piplus\to \kzerob\kplus}=
\sqrt2\epsilon\, \sqrt{{s-4\mkd\over s}}\,
{\sqrt{\lambda_{\eta\pi}(s)}\over 12 F_0^2}\ .
\ena
while the $l=1$ projection of $T_{\eta\piplus\to\eta\piplus}$
vanishes. Using also that the vector form factors $F_V^\pi=F_V^K=1$ at
$O(p^2)$, one easily finds from the unitarity relations of
sec.~\sect{unitfplus}  
\be
\im\fplus(s)=\epsilon {(s-4\mpid)\over 6F_0^2}\im\bar{J}_{\pi\pi}(s)
+\epsilon{(s-4\mkd)\over 12 F_0^2}\im \bar{J}_{KK}(s)\ .
\en
Analogously, the unitarity relation for $\fzero$ involves the $l=0$
partial-wave projections 
\bea
&& {1\over2}\int_{-1}^{1} dz \, T_{\eta\piplus\to \pi^0\piplus}=
\epsilon{ 3s-4\mkd\over 6F_0^2}
\nonumber\\
&& {1\over2}\int_{-1}^{1} dz \, T_{\eta\piplus\to \eta\piplus}=
{\mpid\over 3 F_0^2}\nonumber\\
&& {1\over2}\int_{-1}^{1} dz \, T_{\eta\piplus\to \kzerob\kplus}=
\sqrt6\,{3s-4\mkd\over 12 F_0^2}\ .
\ena
Using the unitarity relations in sec.~\sect{unitfzero} and the
$O(p^2)$ expressions for the scalar form factors
$f_0^{\pi\pi}=f_0^{KK}=1$ and $f_0^{\eta\pi}=\epsilon$, one obtains
\bea
&& \im\fzero(s)=-\epsilon {\Delta_{\pizero\piplus}\over\Delta_{\eta\pi^+}}
{(3s-4\mkd)\over6F_0^2}\im\bar{J}_{\pi\pi}(s)
\nonumber\\
&&\phantom{\im\fzero(s)}
+\epsilon{\mpid\over3F_0^2}\im\bar{J}_{\eta\pi}(s)
+{\Delta_{K^0K^+}\over\Delta_{\eta\pi}}{\sqrt3(3s-4\mkd)\over12F_0^2}
\im\bar{J}_{KK}(s)\ . 
\ena
Dropping the double isospin-suppressed term and
expanding $\Delta_{K^0K^+}=m^2_{K^0}-m^2_{K^+}$ at
$O(p^2)$ and $O(e^2)$ one recovers exactly the
imaginary part of the $O(p^4)$ formula~\rf{fzerop4}.

\section{Dalitz plot parameters}\lblsec{dalitzparam}
For the charged decay, $\eta\to \pi^+\pi^-\pi^0$, a point inside the
Dalitz plot may be  determined in terms of two coordinates $X$,$Y$ defined as
\be
X=\sqrt3 {T_{\piplus}-T_{\piminus}\over Q_c},\quad Y=3{T_{\pizero}\over Q_c}-1
\en 
where $T_{\pi_i}=p^0_{\pi_i}-m_{\pi_i}$ is the kinetic energy of the
pion $\pi_i$ in the $\eta$ rest frame and 
$Q_c=\sum_i T_{\pi_i}=\meta-2\mpiplus-\mpizero$. 
In terms of the Mandelstam variables, one has
\be
X={\sqrt3(u-t)\over 2\meta Q_c},\quad
Y={3( (\meta-\mpizero)^2-s)\over 2\meta Q_c}-1\ .
\en
The Dalitz plot coefficients parametrise the variation of the square
of the amplitude from the center of the  plot
\be
\rho_c(X,Y)=
{\vert{\cal M}_c(X,Y)\vert^2\over\vert{\cal M}_c(0,0)\vert^2}
= 1+aY+bY^2+dX^2+fY^3+GX^2Y+\cdots\ .
\en
The parametrisation accounts for the invariance of the amplitude under
the transformation $X\to -X$ which results from charge conjugation
invariance. 

In the case of the decays into three neutral pions one similarly
introduces two variables 
\be\lbl{xyvars0}
X=\sqrt3{T_{\pi^0_1}-T_{\pi^0_2}\over Q_n},\quad Y=3{T_{\pi^0_3}\over Q_n}-1
\en 
with $Q_n=\meta-3\mpizero$. The amplitude is  invariant under Bose
symmetry transformations $\pi^0_i\leftrightarrow \pi^0_j$. Using
eq.~\rf{xyvars0}, one deduces that it must be invariant 
under the following transformations of the $X$, $Y$ variables
\be
X\to -X,\quad (X+iY)\to \exp({-i\pi\over3})(X-iY)\ .
\en
The expansion of the amplitude squared around the center of the Dalitz
plot thus has the form,
\be
\rho_n(X,Y)=
{\vert{\cal M}_n(X,Y)\vert^2\over \vert{\cal M}_n(0,0)\vert^2}=
 1+2\alpha(X^2+Y^2)+2\gamma(3X^2Y-Y^3)+\cdots \ .
\en
\section{Second-class amplitude in $\tau\to\pi^0\pi^+\nu$
  decay}\lblsec{taupipi2nd} 
We remark here that the $\pi\pi$ scalar form factor $f_0^{\pi\pi}$
(see eq.~\rf{f0pipi}), while involving no resonance contribution (to
first order in isospin breaking) can be estimated in an essentially
model independent way in the low-energy region. The
$\pi^0\pi^+$ system with $l=0$ must be in an isospin $I=2$ state. 
From Watson's theorem, the phase of the scalar form factor $\phi_0$,
must coincide with the $l=0$, $I=2$ $\pi\pi$ scattering phase shift
$\delta_0^2$ in an energy range $s  < s_{in}\simeq 1\ \hbox{GeV}^2$
where  $\pi\pi$ scattering is elastic to a good approximation. We can
then express $f_0^{\pi\pi}$ as a phase dispersive representation,
\be\lbl{f0pipiomnes}
f_0^{\pi\pi}(s)=\exp\left({s\over\pi}
\int_{4\mpid}^\infty ds'{\delta_0^2(s')\over s'(s'-s)}\right)
\exp\left({s\over\pi}
\int_{s_{in}}^\infty ds'{\phi_0(s')-\delta_0^2(s')\over
  s'(s'-s)}\right)\  
\en
(using that $f_0^{\pi\pi}(0)=F_V^\pi(0)=1$).
At energies $s << s_{in}$ we can neglect the effect of the second
exponential in eq.~\rf{f0pipiomnes} and thus obtain an approximation
of the form factor in terms of the known $I=2$ phase shift.

%%%%%%%%%%%%%%%%%%%
\begin{figure}[hbt]
\bc
\includegraphics[width=0.7\linewidth]{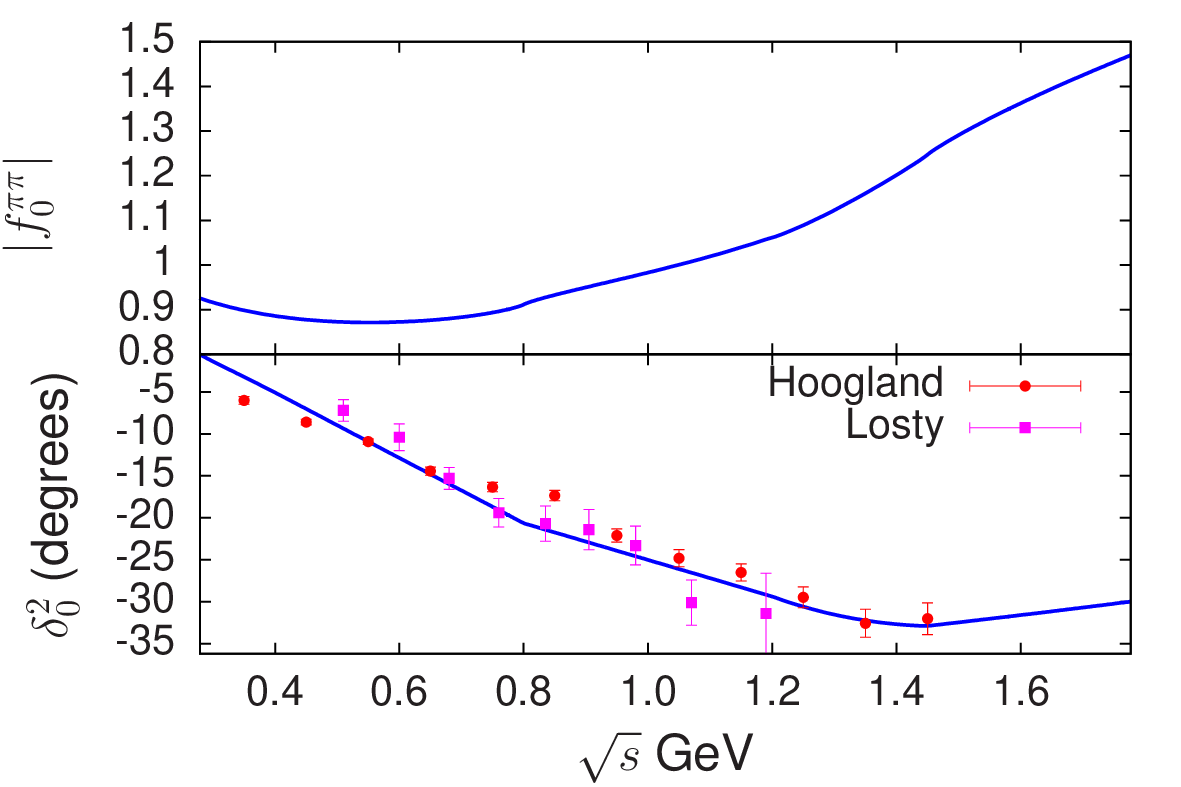}\\
\includegraphics[width=0.7\linewidth]{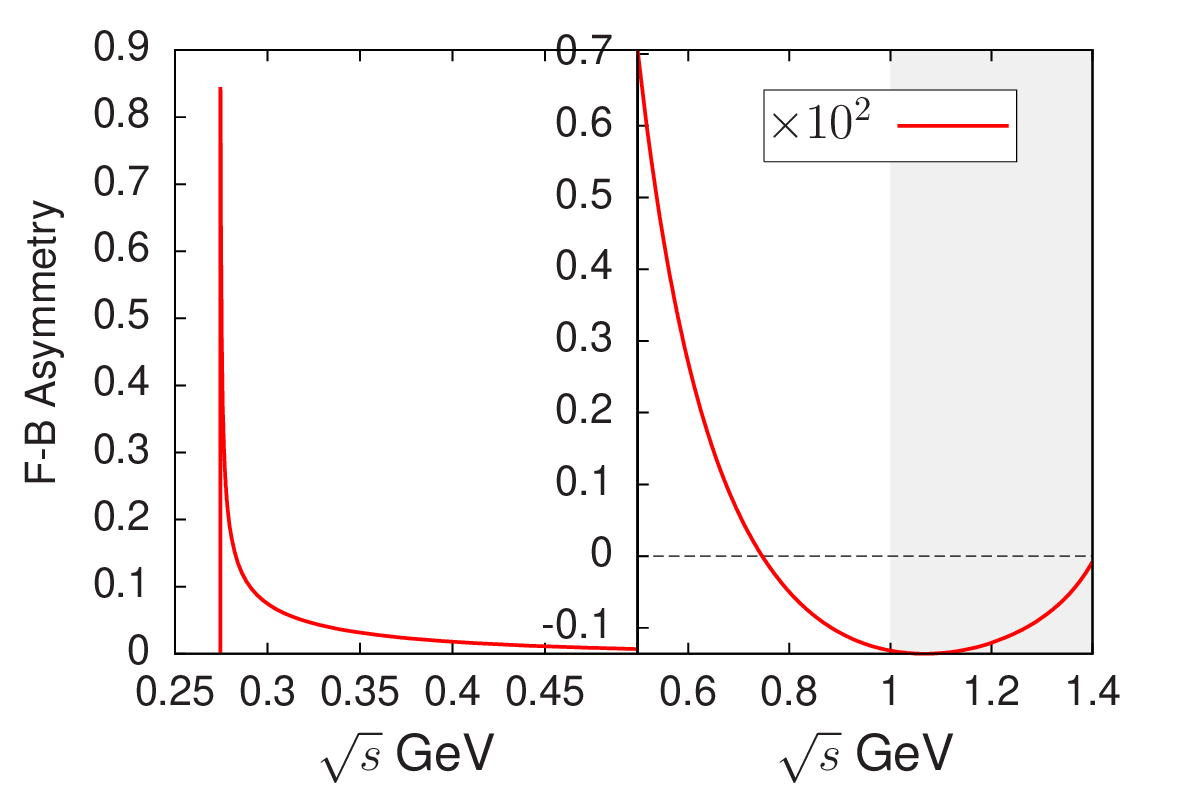}
\caption{\small The upper figure shows the $I=2$ $S$-wave  $\pi\pi$
  phase shift (the experimental data are taken from
  refs.~\cite{Losty:1973et,Hoogland:1977kt}) and 
the scalar form factor from eq.~\rf{f0pipiomnes}, neglecting the second
exponential. The lower figure shows the forward-backward asymmetry in
the $\tau\to \pi^0\pi^+\nu$ decay as a function of the $\pi\pi$
energy. The shaded area indicates the region where the calculation
becomes unreliable.}
\label{fig:f0pipi}
\ec
\end{figure}
%%%%%%%%%%%%%%%%%%%
The form factor $f_0^{\pi\pi}$ plays a role in the search for CP
violation~\cite{Avery:2001md}. Let us consider here its effect in
generating a forward-backward asymmetry in the
$\tau\to\pi^0\pi^+\nu$ decay, 
\be
{\cal A}_{\pi\pi}(s)=
\dfrac{\int_0^1 d\cos\theta {d^2\Gamma\over ds d\cos\theta}-
\int_{-1}^0 d\cos\theta {d^2\Gamma\over ds d\cos\theta}}
{\int_0^1 d\cos\theta {d^2\Gamma\over ds d\cos\theta}+
\int_{-1}^0 d\cos\theta {d^2\Gamma\over ds d\cos\theta}}
\en
where $\theta$ is the angle between the three-momenta of the $\pi^+$
and the $\tau$ in the $\pi\pi$ center-of-mass system. In the energy
range $ s < s_{in}$ one  can express the FB asymmetry
in terms of the moduli of the form factors and the $I=1,2$
phase shifts,
\be
{\cal A}_{\pi\pi}(s)=
\dfrac{3\Delta_{\piplus\pizero}\sqrt{\lambda_{\piplus\pizero}(s)}\vert
  F_V^\pi(s)\vert\vert f_0^{\pi\pi}(s)\vert\cos(\delta_1^1-\delta_0^2)}
{\vert F_V^\pi(s)\vert^2\lambda_{\piplus\pizero}(s)(1+{2s/ \mtaud})
+3\vert f_0^{\pi\pi}(s)\vert^2 \Delta^2_{\piplus\pizero} }\ .
\en
Fig.~\fig{f0pipi} shows that the asymmetry is very small except however
in the energy region $\sqrt{s}\le 300$ MeV where it is positive and larger that
10\%.

\newpage
\bibliography{essai}
\bibliographystyle{epj}
\end{document}

%% file: newcommands.tex
\newcommand\TT{\rule{0pt}{2.5ex}}        %top strut// espaces dans les tables
\newcommand{\be}{\begin{equation}}
\newcommand{\en}{\end{equation}}
\newcommand{\bea}{\begin{eqnarray}}
\newcommand{\ena}{\end{eqnarray}}
\newcommand{\lbl}[1]{\label{eq:#1}}
\newcommand{\lbltab}[1]{\label{tab:#1}}
\newcommand{\lblfig}[1]{\label{fig:#1}}
\newcommand{\lblsec}[1]{\label{sec:#1}}
\newcommand{\rf}[1]{(\ref{eq:#1})}
\newcommand{\Table}[1]{\ref{tab:#1}}
\newcommand{\fig}[1]{\ref{fig:#1}}
\newcommand{\sect}[1]{\ref{sec:#1}}
\newcommand{\braque}[1]{{\langle #1 \rangle}}
\newcommand{\ket}[1]{{\vert #1\rangle}}
\newcommand{\outleft}{%
\mathrel{\setbox1=\hbox{$\scriptstyle{out}$}\setbox0=\hbox{$\langle$}\copy0\kern-0.3\wd0\lower1.1\ht0\copy1\kern-0.4\wd1}}
\newcommand{\outbraque}[1]{{\outleft #1 \rangle}}
\newcommand{\bc}{\begin{center}}
\newcommand{\ec}{\end{center}}
\newcommand{\bt}{\begin{tabular}}
\newcommand{\et}{\end{tabular}}
\newcommand{\ba}{\begin{array}}
\newcommand{\ea}{\end{array}}
\newcommand{\gapprox}{%
\mathrel{%
\setbox0=\hbox{$>$}\raise0.6ex\copy0\kern-\wd0\lower0.65ex\hbox{$\sim$}}}
\newcommand{\lapprox}{%
\mathrel{%
\setbox0=\hbox{$<$}\raise0.6ex\copy0\kern-\wd0\lower0.65ex\hbox{$\sim$}}}
\renewcommand{\slash}[1]{%
\mathrel{\setbox0=\hbox{$/$}\copy0\kern-\wd0\hbox{$#1$}}}
\newcommand{\im}{{\rm Im\,}}
\newcommand{\re}{{\rm Re\,}}
\newcommand{\mpd}{m_P^2}
\newcommand{\mud}{\mu^2}
\newcommand{\meta}{m_\eta}
\newcommand{\barmeta}{\bar{m}_\eta}
\newcommand{\barmetad}{\bar{m}_\eta^2}
\newcommand{\metad}{m_\eta^2}

\newcommand{\mpi}{m_\pi}
\newcommand{\mpid}{m_\pi^2}
\newcommand{\fpid}{F_\pi^2}
\newcommand{\mkd}{m_K^2}

\newcommand{\mtaud}{m_\tau^2}
\newcommand{\fplus}{f^{\eta\pi}_+}
\newcommand{\dotfplus}{\dot{f}^{\eta\pi}_+}
\newcommand{\fzero}{f^{\eta\pi}_0}
\newcommand{\dotfzero}{\dot{f}^{\eta\pi}_0}

\newcommand{\KzeroK}{{\bar{K}^0 K^+}}
\newcommand{\kzerob}{{\bar{K}^0}}
\newcommand{\mpiplus}{m_{\pi^+}}
\newcommand{\mpizero}{m_{\pi^0}}

\newcommand{\mkplusd}{m_{K^+}^2}
\newcommand{\mkzerod}{m_{K^0}^2}
\newcommand{\kzero}{{K^0}}
\newcommand{\kplus}{{K^+}}
\newcommand{\piplus}{{\pi^+} }
\newcommand{\piminus}{{\pi^-} }
\newcommand{\pizero}{{\pi^0} }
\newcommand{\Vmud}{j_\mu^{ud}}

\newcommand{\jbar}{\bar{J}}
\newcommand{\disc}{\hbox{disc}}
\newcommand{\mmd}{m_-^2}
\newcommand{\Kbar}{\bar{K}}
\newcommand{\Gevd}{{$\hbox{GeV}^2\ $}}
\newcommand{\figwidth}{0.7\linewidth}

\makeatletter
\newcommand*\if@single[3]{%
  \setbox0\hbox{${\mathaccent"0362{#1}}^H$}%
  \setbox2\hbox{${\mathaccent"0362{\kern0pt#1}}^H$}%
  \ifdim\ht0=\ht2 #3\else #2\fi
  }
\newcommand*\rel@kern[1]{\kern#1\dimexpr\macc@kerna}
\newcommand*\widebar[1]{\@ifnextchar^{{\wide@bar{#1}{0}}}{\wide@bar{#1}{1}}}
\newcommand*\wide@bar[2]{\if@single{#1}{\wide@bar@{#1}{#2}{1}}{\wide@bar@{#1}{#2}{2}}}
\newcommand*\wide@bar@[3]{%
  \begingroup
  \def\mathaccent##1##2{%
    \if#32 \let\macc@nucleus\first@char \fi
    \setbox\z@\hbox{$\macc@style{\macc@nucleus}_{}$}%
    \setbox\tw@\hbox{$\macc@style{\macc@nucleus}{}_{}$}%
    \dimen@\wd\tw@
    \advance\dimen@-\wd\z@
    \divide\dimen@ 3
    \@tempdima\wd\tw@
    \advance\@tempdima-\scriptspace
    \divide\@tempdima 10
    \advance\dimen@-\@tempdima
    \ifdim\dimen@>\z@ \dimen@0pt\fi
    \rel@kern{0.6}\kern-\dimen@
    \if#31
      \overline{\rel@kern{-0.6}\kern\dimen@\macc@nucleus\rel@kern{0.4}\kern\dimen@}%
      \advance\dimen@0.4\dimexpr\macc@kerna
      \let\final@kern#2%
      \ifdim\dimen@<\z@ \let\final@kern1\fi
      \if\final@kern1 \kern-\dimen@\fi
    \else
      \overline{\rel@kern{-0.6}\kern\dimen@#1}%
    \fi
  }%
  \macc@depth\@ne
  \let\math@bgroup\@empty \let\math@egroup\macc@set@skewchar
  \mathsurround\z@ \frozen@everymath{\mathgroup\macc@group\relax}%
  \macc@set@skewchar\relax
  \let\mathaccentV\macc@nested@a
  \if#31
    \macc@nested@a\relax111{#1}%
  \else
    \def\gobble@till@marker##1\endmarker{}%
    \futurelet\first@char\gobble@till@marker#1\endmarker
    \ifcat\noexpand\first@char A\else
      \def\first@char{}%
    \fi
    \macc@nested@a\relax111{\first@char}%
  \fi
  \endgroup
}
\makeatother